\documentclass[10pt, conference, compsocconf]{IEEEtran}
\usepackage{listings}
\usepackage[font=small,skip=\baselineskip]{caption}
\lstnewenvironment{mycode}[1][]
{
  \minipage[b]{0.3\textwidth}
  \lstset{
    captionpos=b,
    language=C,
    numbers=left,
    basicstyle=\tiny\ttfamily,
    columns=fullflexible,
    showstringspaces=false,
    numbersep=3pt,
    #1
  }
}
{
  \endminipage
}
\usepackage{mathptmx} 
\usepackage{comment}
\usepackage{fancyhdr}
\usepackage[normalem]{ulem}
\usepackage[hyphens]{url}
\usepackage[sort,nocompress]{cite}
\usepackage[final]{microtype}
\usepackage{flushend}
\usepackage[bookmarks=true,breaklinks=true,letterpaper=true,colorlinks,linkcolor=black,citecolor=blue,urlcolor=black]{hyperref}
\usepackage{fancyhdr}
\usepackage[displaymath,mathlines,pagewise]{lineno}
\usepackage{color,soul}
\definecolor{lightgreen}{RGB}{195, 233, 211}
\sethlcolor{lightgreen}
\usepackage[ruled,vlined,linesnumbered]{algorithm2e}
\usepackage{algorithmic}

\usepackage{cite}
\usepackage{indentfirst}
\usepackage{setspace}
\usepackage{mathdesign}
\usepackage{adjustbox}
\usepackage{graphicx,dblfloatfix}
\usepackage{caption}
\usepackage{epstopdf}
\usepackage[nopar]{lipsum}
\usepackage{graphicx}
\usepackage{mathtools}
\usepackage{amssymb,amsmath}
\usepackage[numbers,compress]{natbib}
\usepackage{footnote}
\makesavenoteenv{tabular}
\makesavenoteenv{table}
\usepackage{tablefootnote}
\usepackage[labelfont=bf]{caption}
\usepackage{array}
\usepackage{wasysym}
\usepackage{multirow}
\usepackage{xfrac}
\usepackage[utf8]{inputenc}
\usepackage{etoolbox}
\ifCLASSINFOpdf
\else
\fi
\newbool{inccomment}
\setbool{inccomment}{true}
\newcommand{\sms}[1]{\ifbool{inccomment}{{\color{blue}#1}}{}}
\newcommand{\akj}[1]{\ifbool{inccomment}{{\color{Green}AKJ #1}}{}}
\newcommand{\rami}[1]{\ifbool{inccomment}{{\color{red}rami #1}}{}}
\usepackage{comment}

\begin{document}
%
\pagenumbering{arabic}
\title{\LARGE \textbf{Mitigating Wordline Crosstalk using Adaptive Trees of Counters}}

\newcommand{\superscript}[1]{\ensuremath{^{\textrm{#1}}}}
\def\sharedaffiliation{\end{tabular}\newline\begin{tabular}{c}}
\def\wg{\superscript{\dag}}
\def\wu{\superscript{\S}}

\author{\IEEEauthorblockN{ Seyed Mohammad Seyedzadeh\wg, Alex K. Jones\wu, Rami Melhem\wg}
\IEEEauthorblockA{Computer Science Department\wg, Electrical and Computer Engineering Department\wu\\
University of Pittsburgh\\
seyedzadeh@cs.pitt.edu, akjones@pitt.edu, melhem@cs.pitt.edu\vspace{-0.2em}}
}


%
\maketitle

\thispagestyle{plain}
\pagestyle{plain}
\begin{abstract}

DRAM technology scaling has the undesirable side effect of degrading cell reliability. One such concern of deeply scaled DRAMs is the increased coupling between adjacent cells, commonly referred to as crosstalk. High access frequency of certain rows in the DRAM may cause data loss in cells of physically adjacent rows due to crosstalk. The malicious exploit of this crosstalk by repeatedly accessing a row to induce this effect is known as \textit{row hammering}. Additionally, inadvertent row hammering may also occur due to the natural weighted nature of applications' access patterns.

In this paper, we analyze the efficiency of existing approaches for mitigating wordline crosstalk and demonstrate that they have been conservatively designed. Given the unbalanced nature of DRAM accesses, a small group of dynamically allocated counters in banks can deterministically detect ``hot'' rows and mitigate crosstalk. Based on our findings, we propose a Counter-based Adaptive Tree (CAT) approach to mitigate wordline crosstalk using adaptive trees of counters to guide appropriate refreshing of vulnerable rows. The key idea is to tune the distribution of the counters to the rows in a bank based on the memory reference patterns. In contrast to deterministic solutions, CAT utilizes fewer counters, making it practically feasible to be implemented on-chip. Compared to existing probabilistic approaches, CAT more precisely refreshes rows vulnerable to crosstalk based on their access frequency.

Experimental results on workloads from four benchmark suites show that
CAT reduces the Crosstalk Mitigation Refresh Power Overhead in quad-core systems to 7\%, which is an improvement over the 21\% and 18\% incurred in the leading deterministic and probabilistic approaches, respectively. Moreover, CAT incurs very low performance overhead ($\sim 0.5\%$). Hardware synthesis evaluation shows that CAT can be implemented on-chip with only a nominal area overhead.

\end{abstract}

\IEEEpeerreviewmaketitle

\section{\textbf{INTRODUCTION}}

Dynamic random-access memory (DRAM) has been a widely used storage element in computer systems. Over time, process technology has scaled DRAM to achieve greater storage density by reducing the technology feature size~\cite{volos2016effective,mueller2005challenges,zhang2016restore,sinha2012exploring,zeng2011memscale}. However, shrinking process technology causes DRAM cells to become significantly less reliable~\cite{mukundan2013understanding,kim2005technology, mueller2005challenges}. As the chip density increases with technology scaling, the interaction between circuit components, such as transistors, capacitors, and wires increases, leading voltage fluctuations~\cite{cha2011dram,mandelman2002challenges,
konishi1989analysis}.
Specifically, when the cumulative interference to a DRAM wordline becomes strong enough, the state of nearby cells can change leading to memory errors~\cite{park2016experiments,park2014active}. 
Vulnerability to wordline electromagnetic coupling (crosstalk) exists in recent sub 40nm commodity DRAM chips due to physical limitations of these technologies. Kim~\cite{kim2014flipping} showed that through frequently alternating the charge of specific memory locations, crosstalk can be used, intentionally, to affect the charge of adjacent cells~\cite{gruss2015rowhammer}. However, in addition to intentional malicious attacks~\cite{aweke2016anvil,armor}, inadvertent row hammering is also possible due to the unbalanced nature of some applications' access patterns that can lead to ``hot'' rows and induce crosstalk.

One simple solution to mitigate wordline crosstalk is to increase the refresh rate for all rows. Although, this approach is effective, it imposes an unnecessarily high power and performance overhead~\cite{aweke2016anvil,rahmati2014refreshing,mukundan2013understanding,kotra2017hardware,kim2003block,nair2013case,ohsawa1998optimizing,arjomand2016boosting,liu2012flikker}. There are two main hardware approaches to mitigate wordline crosstalk in  DRAM. The first relies on a random number generator \cite{yang201416,srinivasan20102} in the memory controller to probabilistically refresh accessed rows \cite{kim2015architectural}.
Although the idea behind this approach is simple, it results in an early refresh of  rows that can be potentially affected by crosstalk~\cite{aweke2016anvil}, as well as unnecessary row refreshing in the absence of hot rows. 
    
 The second approach detects the most frequently accessed rows, or \emph{aggressor rows}, and then refresh the rows that are adjacent to it, or \emph{victim rows}.  To protect new DDR4 modules against wordline crosstalk, DRAM architectures implement the Targeted Row Refresh~(TRR) mechanism that allows victim rows to be refreshed on demand\footnote{Note that Micron DDR4 documentation~\cite{micron2} suggests that the TRR module is an optional module, thus there may exist future DDR4 systems that do not incorporate TRR.}.  Intel has announced the existence of pseudo-targeted row refresh in Xeon-class Ivybridge architectures to help in the crosstalk mitigation, but has not yet released the details of this mechanism. Several methods have been proposed to identify rows for TRR. A simple method to deterministically recognize aggressor rows, called Static Counter Assignment (SCA), is to dedicate a counter per row to keep track of the number of row activations. However, having one counter per row induces a significant area and power overhead to the memory system. To prevent this high cost, row activation counters can be stored in a reserved area of the main memory and a dedicated on-chip counter-cache can be used to minimize the performance penalty of retrieving counter values from main memory\cite{kim2015architectural}.

	Due to row access locality in DRAM~\cite{jeong2012balancing,2016power,agrawal2014mosaic}, many counters in SCA would be underutilized~\cite{seyedzadeh2017counter}. Thus, we propose a Counter-based Adaptive Tree (CAT) approach that dynamically assigns counters to frequently accessed aggressor rows. Hence, with a small number of on-chip counters it is possible to deterministically refresh victim rows. When CAT results in a highly unbalanced tree, it provides a significant advantage in refresh energy over a block-based uniform counter distribution with a similar number of counters, while CAT also converges to a balanced tree when accesses in memory are uniform. 
    The dynamic counter tree of CAT is constructed by partitioning the memory rows into groups, each governed by a counter, based on the DRAM access pattern. To accomplish this, the \textbf{refresh threshold}, defined as the number of aggressor row accesses before crosstalk is known to occur in victim rows, is subdivided into different split thresholds. These split thresholds identify points at which a group is likely to contain an aggressor row and increased precision is desired for more precise identification.  
    When crossing a split threshold, CAT subdivides the block governed by a counter and provides a second counter to increase the precision. Thus, CAT builds a uniform tree for workloads with random row access frequency and a non-uniform tree for workloads with biased row access frequency. The key feature of CAT is that hot rows are instrumented using smaller groups, while rows with low access frequency (cold rows) are unlikely to induce crosstalk and are instrumented using larger groups.  
    Thus, the selection of appropriate split thresholds is a key factor to the success of CAT. 
    
CAT deterministically detects aggressor rows before they reach the refresh threshold, whether they result from intentional malicious attacks or applications' unbalanced access pasterns. It refreshes victim rows to prevent crosstalk and guarantee reliability of the memory during normal operation as well as defend  against row hammering attacks. We make the following contributions for wordline crosstalk mitigation:
\begin{itemize}
\item We demonstrate that, due to access locality in DRAM~\cite{jeong2012balancing,2016power,agrawal2014mosaic}, instead of over-provisioning with one counter per row,   a small number of counters can be implemented on-chip to refresh victim rows, while achieving low latency, low power consumption and satisfactory area overhead.
\item We introduce a non-uniform counter assignment, CAT, to more precisely determine the aggressor rows and refresh vulnerable rows to improve the effectiveness of uniform counter assignment.
\item We determine the suitable number and values of split thresholds for non-uniform counter assignment (\textit{i.e.,} CAT) to best construct a, potentially unbalanced, tree for optimized alignment of access counters to increasingly small groups of rows that contain aggressor rows.
\item We introduce a dynamically reconfigurable CAT scheme (DRCAT), that tracks and reacts to the temporal changes in memory access patterns resulting from either application context switching or different phases of a particular application in order to more precisely identify actual victim rows and reduce DRAM refresh energy.
\end{itemize}

\section{\textbf{BACKGROUND AND RELATED WORK}}

DRAM-based main memory is a multi-level hierarchy of structures. Each memory module is composed of a number of chips and is connected to the memory controller through a channel. Internally, each chip consists of multiple banks and each bank is organized as rows of DRAM cells. Each cell is composed of an access transistor and a capacitor, in which the data is stored. While accessing a row, all cells in the row are selected in parallel using a wordline. Due to capacitive coupling between cells on adjacent wordlines, if a wordline (aggressor row) is accessed frequently, voltage levels on neighboring wordlines (victim rows) can be affected leading to crosstalk. 
Mitigating wordline crosstalk is possible by refreshing the victim rows before the aggressor rows reach the refresh threshold.

A hardware approach to alleviate wordline crosstalk is for each DRAM access to refresh the victim rows adjacent to the accessed row based on a probability function~\cite{kim2015architectural}. In this probabilistic approach, called PRA (Probabilistic Row Activation), the memory controller uses a Pseudo-Random Number Generator with a given probability~($\alpha$) to determine when the memory controller should issue a refresh signal to refresh the two rows adjacent to the accessed row. When either the number of memory accesses or the probability $\alpha$ is high, this approach generates a significant number of refresh commands, thus exacerbating memory contention and increasing the energy cost~\cite{aweke2016anvil}.

As a hardware alternative to the probabilistic approach, a deterministic approach can be used to prevent aggressor rows from being accessed more than the refresh threshold before refreshing the victim rows. Maintaining a counter for each memory row is a significant overhead~\cite{bains2016distributed,greenfield2015row}. To address this problem, an approach was proposed that stores the counters in a reserved area of DRAM and a set-associative counter cache was established in the memory controller to improve accessibility to frequently used counters~\cite{kim2015architectural}. Note that the primary idea in \cite{kim2015architectural} is similar to that used for Counter-based caches~\cite{kharbutli2005counter}, where threshold-based counters detect expired lines for proactive eviction. While, using counters allows for accurate counts of row accesses, caching the counters introduces the complexity of maintaining a cache (e.g., tag matching, eviction policies) within the memory controller. Moreover, misses to the cache counter can be expensive.

Rewriting instructions, such as CLFLUSH~\cite{seaborn2015exploiting}, have been proposed as software countermeasures against wordline crosstalk and are now deployed in Google Native Client. Similarly, access to the Linux pagemap interface is now prohibited from userland~\cite{shutemovpagemap}. These countermeasures have already been proven insufficient to mitigate malicious kernel attacks \cite{bosman2016dedup,gruss2016rowhammer}. In \cite{aweke2016anvil}, a generic software mechanism, ANVIL, is proposed to detect aggressor rows by monitoring the last-level cache miss rate and row accesses with high temporal locality. A similar approach is proposed in \cite{herath2015these} to monitor the number of last-level cache misses during a given refresh interval. Both approaches rely on software access to CPU performance counters.


\section{\textbf{MOTIVATION}}\label{motivation}

In this section, we analyse the previously proposed hardware approaches and make key observations to motivate the dynamic counter assignment as a hardware solution that mitigates wordline crosstalk and combats row hammering.
\subsection{Probabilistic Refresh Analysis}
Using a probabilistic approach, such as PRA, to mitigate wordline crosstalk can protect against failure with a high probability, depending on the value of the refresh threshold, $T$, and the probability, $\alpha$, of triggering a refresh. The probability of experiencing an error in Y years (defined as Y-years unsurvivability) for PRA is computed as:
\begin{equation}\label{prob_pra}
unsurvivability = (1-p)^{T} \times Q_{0} Q_{1}
\end{equation}
where $p = \alpha$ is the probability of refreshing TWO victim rows on an access, $Q_{0}$ is the number of refresh threshold windows during a refresh interval, and $Q_{1}$ is the number of 64ms periods during Y years. The parameter $T$ depends on the technology node. Specifically, scaling down DRAM increases voltage fluctuations in cells because of the interaction between circuit components. Therefore, the refresh threshold is projected to decrease for future memory technology~\cite{kim2015architectural}. 

Figure~\ref{pra_reliability} compares the 5-years unsurvivability for different refresh thresholds when $p$ ranges from 0.001 to 0.006. Assuming mild row accesses during refresh intervals, we set $Q_{0}$ to 10, 15, 20, and 40. Figure~\ref{pra_reliability} shows that for T=32K and $p > 0.001$, PRA's unsurvivability is lower than the Chipkill's unsurvivability of 1E-4. Our key observation from this figure is that, for smaller values of $T$, larger values of $p$ are needed to match the 5-years survivability of chipkill\footnote{Similar analysis done in \cite{kim2015architectural} shows that $PRA_{p=0.001}$ probability of failure is higher than 1E-4.}. In fact, PRA's failure probability increases exponentially when the refresh threshold scales down, as is expected in future technology nodes. This means that larger values of $p$ (more frequent random refreshes) are needed to guarantee acceptable survivability. 

Note that the reliability reported in Figure~\ref{pra_reliability} assumes the use of a true pseudo random number generator, PRNG, such as the one proposed in ~\cite{srinivasan20102}. This is important since the computed reliability is contingent on the randomness of the numbers generated by PRNG. Specifically, the unsurvivability in Eq.~\ref{prob_pra} will not apply if a simpler (less costly in terms of area and power) PRNG is used since the randomness of the generated numbers will not be independent enough. To study the effect of the randomness of the generated numbers, we conducted a Monte-Carlo simulation to estimate the unsurvivability of PRA when a LFSR-based PRNG~\cite{lfsr1,lfsr2} is used. The results show that, using an LFSR-based PRNG largely increases PRA's unsurvivability. For example, for T=16K and p=0.005, PRA's unsurvivability reaches 1E-4 after only 25 refresh intervals. To improve the reliability, a much larger value of $p$ should be used with LFSR-based PRNGs which increases the refresh power and decreases the performance. A similar conclusion was reached in \cite{armor}.
Hence, PRA requires true random number generators, which are known to be complex and to consume relatively large power~\cite{yang201416,srinivasan20102,zhou2008ultra,bucci2003high,bucci2003high1},
to achieve the probabilities shown in Figure~\ref{pra_reliability}.

\begin{figure}[!t]
\begin{center}
\includegraphics[width=0.485\textwidth]{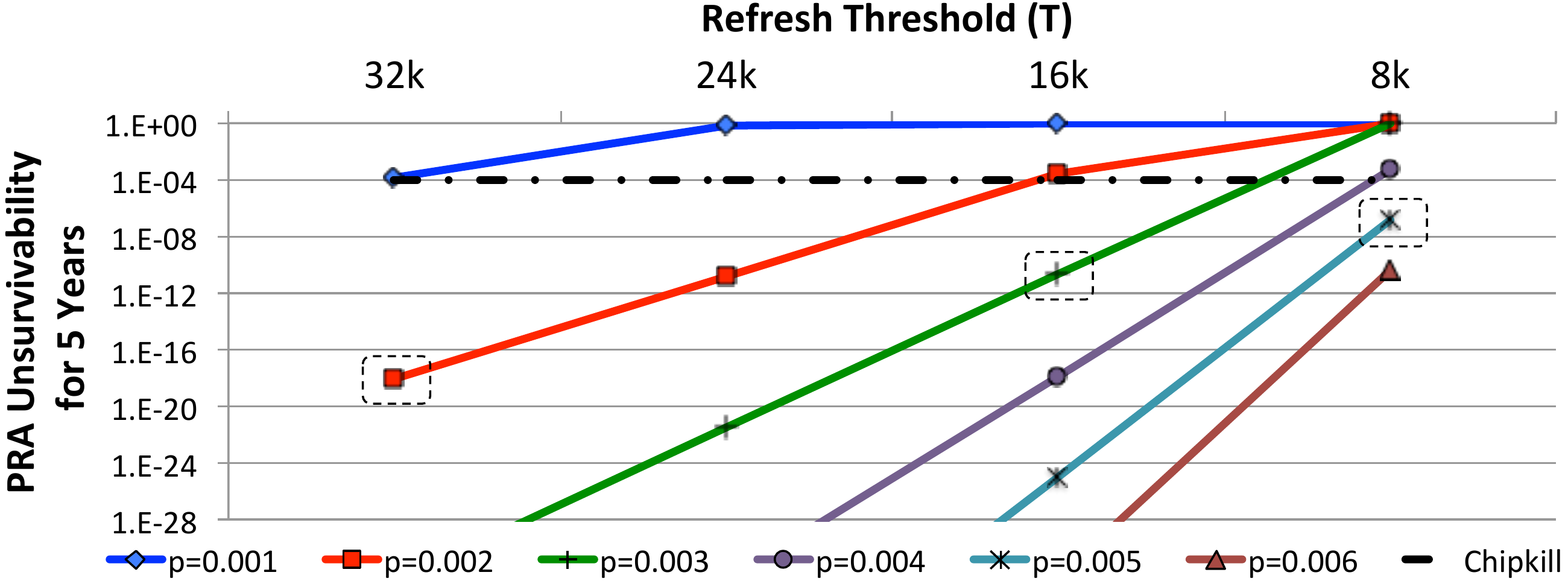}\vspace{-0.5em}
\caption{PRA unsurvivability for refresh thresholds 32k, 24k, 16k and 8k. PRA refreshes two victim rows with  probability $p$.}\label{pra_reliability}
\end{center}\vspace{-2.25em}
\end{figure}

\subsection{Static Counter Assignment~(SCA) Analysis}

Using a deterministic approach for counting the number of accesses per row with on chip counters using SCA requires a large area and power overhead. One intuitive solution is to use fewer counters by partitioning the rows in each memory bank into fixed-size groups and assign one counter per group. To illustrate SCA, we assume that every bank in DRAM includes $N$ rows and uses $M$ counters. The refresh threshold, $T$, determines the size of every counter as $log_{2}T$-bits. This approach, called SCA$_{M}$, divides the rows into $M$ groups, each including $\frac{N}{M}$ rows.
For every row activation, the row address maps to the appropriate counter. Then, the corresponding counter counts the number of accesses. When the counter reaches the threshold, it is reset and a refresh signal is sent to the memory controller to refresh $\frac{N}{M}+2$ rows; the $\frac{N}{M}$ rows in the group plus the two rows adjacent to the group, which guarantees the refresh of any row in or adjacent to the group subjected to the crosstalk. 

The energy overhead in SCA originates from activating the counters when memory is accessed and refreshing $\frac{N}{M}+2$ rows when a counter exceeds $T$. Figure~\ref{fig1007} breaks down the energy overhead of SCA$_{M}$ during a 64$ms$ auto-refresh period when $N=65536$ and the number of counters $M$ ranges from 16 to 65536\footnote{The refresh energy includes the average refresh energy of victim rows for 18 real workloads (Details in Section~\ref{Methodology}). We modified CACTI~\cite{muralimanohar2009cacti} to model the cache in the counter cache approach \cite{kim2015architectural}.}. For a small number of counters, the energy resulting from refreshing victim rows (blue line) dominates the total energy of activating counters in SCA. In contrast, the total energy of activating counters in SCA is the dominant energy as the number of counters significantly increases (orange line). 

Figure~\ref{fig1007} shows that the total energy can be minimized at M=128. In this case, SCA$_{128}$ not only reduces the energy overhead in comparison to SCA$_{65536}$, but also decreases the area overhead by two orders of magnitude (as will be explained in Section~\ref{Evaluation}). 
In comparison, the counter cache approach~\cite{kim2015architectural}, which stores counters in the reserved area of DRAM memory, reports data for a much larger counter storage cache, requiring capacity to store on the order of thousands of counters per bank.  Ostensibly, this is to allow for enough flexibility to store the relevant counters to hot rows without a high miss rate due to capacity misses and/or thrashing. Thus, the energy overhead of counter storage cache will significantly exceed SCA$_{128}$ due to the increased static power.
	For example, Figure~\ref{fig1007} shows the optimistic energy (assuming no misses requiring accesses to the DRAM) of 2K and 8K per bank counter caches as horizontal lines. These lines intersect the SCA$_{4096}$--SCA$_{16384}$ points, respectively, as they have the same amount of total counter storage\footnote{The counter caches have additional storage to store the tag array. However, this storage is typically less than the data array making it inconsequential on a log plot.}. Thus, the total energy consumption of SCA with $M \leq 4K$ counters is lower than that of the counter caches with different sizes.  
In particular, \textit{SCA$_{128}$ can improve the total energy overhead and area overhead by 1.5 orders of magnitude in comparison to a 2K counter cache and nearly two orders of magnitude compared to an 8K counter cache~\cite{kim2015architectural}.} 

Thus, our key observation of these deterministic approaches is that allocating one counter to each row in a DRAM bank with a cache counter can be effective but are somewhat conservative and leave room for improvement.
Specifically, the analysis of row access frequency of DRAM banks on real workloads reveals that the row access frequency during the refresh interval is not uniform and mostly a small group of rows are activated in DRAM banks. For example, Figure~\ref{fig1001} depicts the row access frequency of a given bank for two typical real workloads, \textit{blackscholes} and \textit{facesim}, within a time period of one refresh interval (64$ms$). Figure~\ref{fig1001} clearly shows that a small group of rows dominate overall accesses. \textit{This motives us to propose a dynamic counter assignment for wordline crosstalk mitigation.}

\subsection{Our Goal}

Our goal is to take advantage of a small number of counters per bank, but better target the aggressor rows to provide further benefits to overall energy consumption while mitigating crosstalk. Our novelty is the adaptive construction and dynamic reconfigurability of a "potentially unbalanced" tree of counters to match access patterns.

\begin{figure}[!t]
\begin{center}
\includegraphics[width=0.49\textwidth]{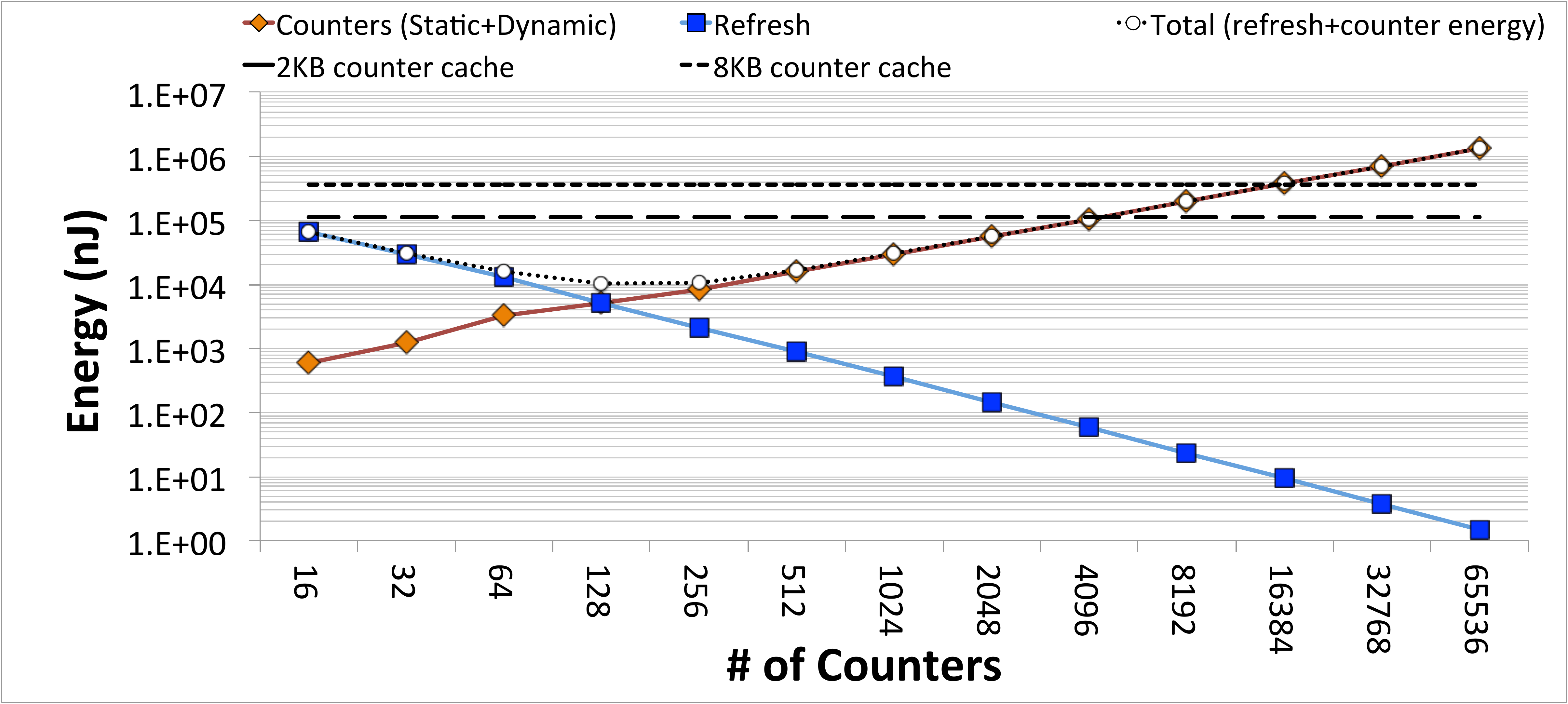}\vspace{-0.5em}
\caption{The energy overhead of SCA and counter caches~\cite{kim2015architectural} for different number of counters.}\label{fig1007}
\end{center}\vspace{-1.5em}
\end{figure}
\begin{figure}[!t]
\begin{center}
\includegraphics[width=0.475\textwidth]{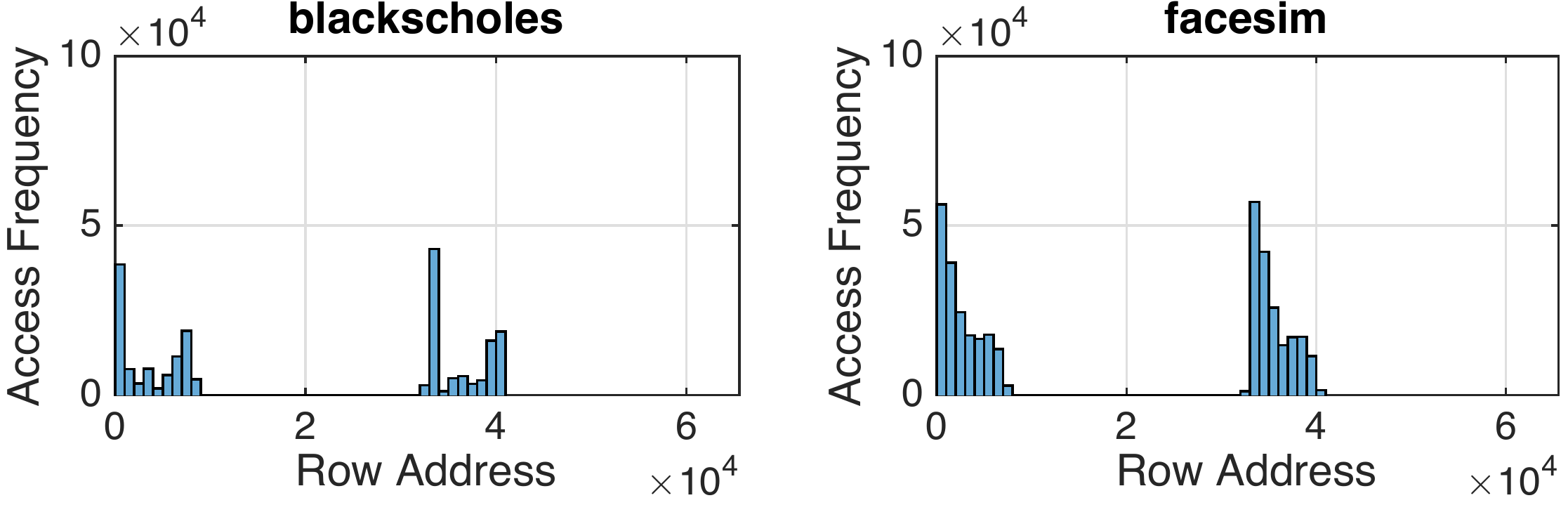}\vspace{-0.75em}
\caption{Row address frequency in a DRAM bank with 64K rows.}\label{fig1001}
\end{center}\vspace{-2em}
\end{figure}

\section{\textbf{COUNTER-BASED ADAPTIVE TREE}}\label{CAT_description}

In order to better assign row partitions to access counters, the Counter-Based Adaptive Tree (CAT) is a new and practical dynamic row partitioning technique that considers access frequency of rows to more carefully assign counters to appropriately sized groups of rows in order to improve energy and area efficiency.
To divide an initial group of rows (e.g., a bank or some other uniform coarse partition) into groups of suitable sizes, CAT defines different split thresholds that identify access frequency stages prior to reaching the refresh threshold. These split thresholds are used to build a non-uniform binary tree structure that maps hot rows to smaller groups, while cold rows, \textit{i.e.} rows with relatively low access frequencies, are mapped to larger groups. This aligns access counters to small groups of rows that contain an aggressor row to more precisely identify actual victim rows. 
\subsection{A simple CAT Example}

Figure~\ref{fig1003} depicts two trees built by CAT, where a terminal node, $\CIRCLE$, represents an active counter and intermediate node, $\Circle$, represents an expired counter, which had been split into two counters. The level of a node is defined as its distance from the root, with the root being at level zero. The levels of the CAT are associated with unique split thresholds. Hence, when a node reaches the next threshold, it further subdivides the group, or \emph{splits} the node, generating two children counters initialized to the current count value. This is accomplished by activating a second counter as a clone of the existing counter. The binary tree of counters continues to grow until all available counters are activated or a maximum allowed level, a parameter of the CAT algorithm, is reached. 

More precisely, assuming that we limit the number of levels in the tree to $L$, we define $L-1$ split thresholds $T_{0},~...,~T_{L-2}$ where $T_{0}\leq...\leq T_{L-1}$ and $T_{L-1}=T$, recalling that $T$ is the refresh threshold.  Each of the $M$ counters in a bank, $C_{0}$,~...,~$C_{M-1}$, has $log_{2}T$ bits and, initially, only $C_0$ is in active mode. When a counter at level $l$ reaches the split threshold, $T_l$, it splits and two counters are activated at level $l+1$. This process continues until all the counters are activated or $l=L-1$. For example, Figure~\ref{fig1003} shows two CATs for $L=6$ and $M=8$. The CAT in Figure~\ref{fig1003}(a) results from a non-uniform row access pattern, which causes more counters to be allocated to the hot row area (smaller blocks) and grows the tree through level 5. In contrast, when the row access frequency is uniform, counters are distributed uniformly throughout the bank addresses as shown in Figure~\ref{fig1003}(b). In this case, the CAT approach grows the tree only through level 3 and mimics SCA.

\begin{figure}[!t]
\begin{center}
\includegraphics[width=0.47\textwidth]{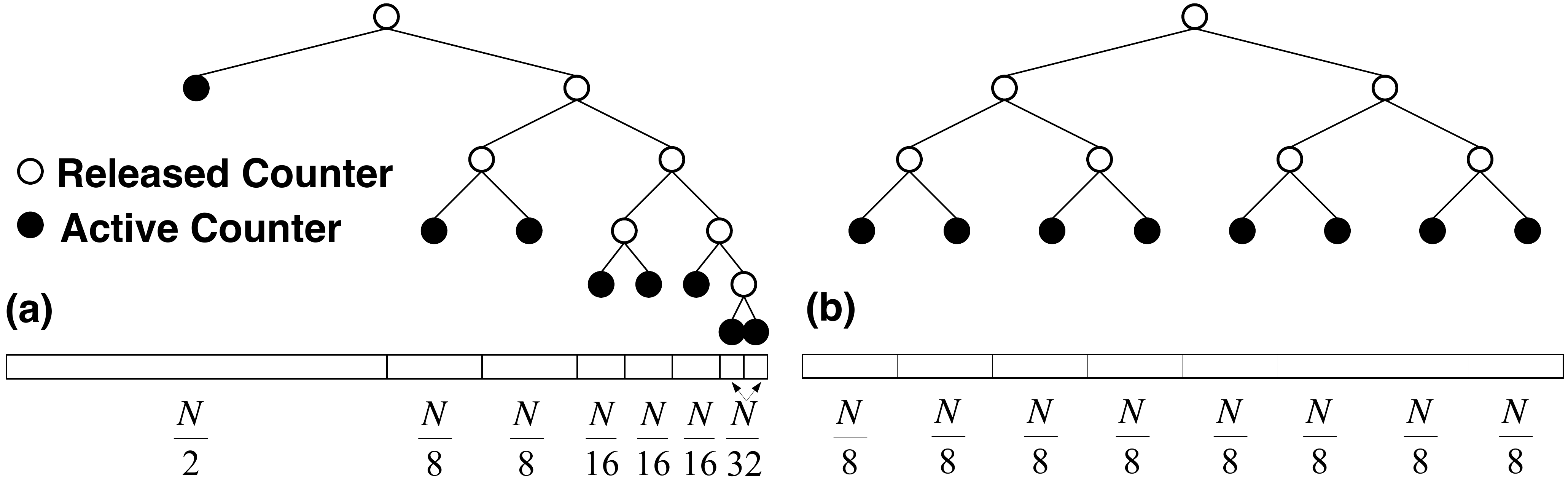}
\caption{The adaptive tress of counters for the workload with (a) biased, (b) uniform row address frequency. The number of row addresses in the bank is $N$.}\label{fig1003}
\end{center}\vspace{-2.5em}
\end{figure}

In CAT, $N$ rows in one bank are initially treated as a group to which $C_{0}$ is allocated. As soon as $C_{0}$ reaches $T_{0}$, CAT splits $C_{0}$ into $C_{0}$ and $C_{1}$ with the same starting value of $T_{0}$. In this case, $C_{0}$ counts the number of accesses when the row address is between 0 to $\frac{N}{2}-1$ and $C_{1}$ counts the number of accesses when the row address ranges from $\frac{N}{2}$ to $N-1$. When $C_{1}$ reaches $T_{1}$, CAT splits $C_{1}$ into $C_{1}$ and $C_{2}$ with the new initial value $T_{1}$ where $C_{1}$ and $C_{2}$ track row addresses in the ranges from $\frac{N}{2}$ to $\frac{3N}{2}-1$ and from $\frac{3N}{2}$ to $(N-1)$, respectively. CAT continues this process until it activates all counters and no group can be split into smaller sub-groups. At this point, the split thresholds of counters are set to $T$. The minimum number of rows in a given group depends on the number of defined split thresholds. With $L-1$ split thresholds (a CAT with at most $L$ levels), the minimum number of rows per group is $\frac{N}{2^{L-1}}$. 

\subsection{Constructing the CAT}

Algorithm~\ref{Tree_algorithm1} shows the process for refreshing rows under the CAT structure per memory bank. It has two main modules: the Counter Module~(CM) that records the number of row accesses and the Reconfiguration Counter Module~(\textit{RCM}) that activates and initializes counters. Assuming $M$ counters in a given bank, CAT requires an array of $M$ counter modules that are implemented on-chip, and one RCM that can be implemented either on-chip or in software. Each counter module $CM_{i}$ maintains two registers, $L_{i}$ and $U_{i}$ to store the lower and upper row addresses assigned to this counter, and a register $l_{i}$ to store the index of the split threshold used for that counter. The RCM maintains a $last\_activated$ counter register.   

Initially, at the start of each refresh interval, CAT is reset such that only the first counter module, $CM_{0}$, is activated with $L_0$= 0, $U_0$ = $N-1$, $l_0$=0, and $last\_activated=0$.  Each time a row is accessed, its address is located in the range $L_{i}$ - $U_{i}$ of some active $C_{i}$, and this counter is incremented (lines~5-7).  When $C_{i}$ reaches $T_{l_{i}}$, $flag_{i}$ is raised (lines~8-10), which triggers RCM to activate a new counter as long as the number of active counters is less than $M$ and the counter level $l_{i}<L-1$~(lines 15-16). When a new counter is activated, it is initialized by $C_{i}$~(line 17) and the interval between $L_{i}$ and $U_{i}$ is split into two equal-size ranges where the lower bound of $C_{i}$ remains unchanged and the upper bound of $C_{i}$ is assigned to the upper bound of the new counter. Then, $U_{i}$ shrinks to $U_{i}=\frac{U_{i}+L_{i}}{2}$ and the lower bound of the new counter is set to $U_{i}+1$ (lines 18-20). The split thresholds of both counters are set to $l_{i}$+1 (lines 21-22).  For example, after initialization, when $CM_0$ reaches $T_{l_0}$, $CM_1$ is introduced by subdividing $CM_0$ in half, such that $C_1=C_0$, $L_0=0$, $U_0=\frac{N}{2}-1$, $L_1 = \frac{N}{2}$, and $U_1=N-1$ with both $CM$'s split thresholds being set to $T_{l_1}$ and $last\_activated=1$.

This process continues until some CM, $CM_{i}$, reaches the highest threshold $T_{l_{i}}=T$ (\textit{i.e.,} if $l_{i}=L-1$, lines 10-12).  In this case, $C_{i}$ is reset and the signal $R_{i}$ is raised to cause the memory controller to refresh all existing rows in the address range of $L_{i}$-1 and $U_{i}$+1.  When all counters are activated, CAT will set the index of all split thresholds to $l_{i}=L-1$ which causes $T_{l_{i}}=T$ (line 25).

\subsection{Efficient CAT Management Using SRAM}

To directly implement Algorithm~\ref{Tree_algorithm1}, maintaining the range boundaries of row blocks requires more storage than the actual counters, themselves. Given that SRAM uses less area and static power than registers~\cite{jacob2010memory}, we are motivated to design and optimize the CAT for SRAM. In this case, instead of storing row range boundaries, we use pointers to \textit{store the structure} of the CAT as shown in Figure~\ref{fig:table_tree}.  During each access, the tree structure is traversed sequentially by chasing the pointers to find the counter assigned to a specific row address. 

The CAT, shown visually in Figure~\ref{fig:table_tree}(a), is composed of two types of nodes: leaf nodes (shown in light blue) that represent active counters and intermediate nodes (shown in white) that determine the tree's structure. Rather than store all the nodes in our data structure, shown in Figure~\ref{fig:table_tree}(b), we store only the intermediate nodes.  Thus, we use an array, $I$, of size $M-1$ (the maximum number of intermediate nodes in a tree with $M$ leaves) to store information about intermediate nodes. Separately, we use another array $C$, of size $M$ to store the counters, shown in Figure~\ref{fig:table_tree}(c).  For each intermediate node, two pointers, $L\_ptr$ and $R\_ptr$, point to information about its two successors.  If the successor is another intermediate node, $L/R\_ptr$ contains the entry for that intermediate node.  If the successor is a leaf, $L/R\_ptr$ contains the entry for the counter corresponding to that leaf. Two flags, $L\_leaf$ and $R\_leaf$, indicate if the corresponding successor is an intermediate or leaf node. The length of each counter is $log(T)$ bits and each pointer is $log(M)$ bits. The root of the tree, $I_0$, is deterministically stored in the first entry of the array $I$. 

\begin{algorithm}[!t]
\footnotesize
\textbf{Parameters:}~\textit{$N:$}~ \#~rows per bank;~$\textit{M:}$~ \#~counters per bank; ~\textit{$L$}:~\#~ thresholds;
~~~~\textbf{Input}:~\textit{$row\_address$};~~~~\textbf{Output}: \textit{$R_{i}$}:~ Refresh signal for refreshing all existing rows between $L_{i}$-1 and $U_{i}$+1.\\
\Begin
{
\textbf{Counter Module CM$_{i}$} /* $i=0,...,M-1$ */\\
\If{$L_{i}~\leq~row\_address~\leq U_{i}$}
{
    \eIf{$C_{i}<T_{l_{i}}$}
    {
    $C_{i}$ ++;
    }{ \eIf{$l_{i}<L-1$}
      {
      $flag_{i}=1$; /* Signal to trigger RCM /*
      }
        {
        $R_{i}$=1; /* Signal to refresh corresponding rows*/\\
        $C_{i}=0$;
        }
        }
    }
\textbf{Reconfiguration Counter Module~(RCM)} /* Activated when $flag_{i}=1$*/\\
\If{$flag_{i}==1$ for some $i$}
		{
				\If{$last\_activated~<~M-1$ \&\& $l_{i}~<~L-1$}
                {
   					$last\_activated$++; /*Increase \# of active counters*/\\
                    $C_{last\_activated}=C_{i}$;\\
					$U_{last\_activated}$=$U_{i}$;\\
                    $U_{i}=\frac{U_{i}+L_{i}}{2}$;\\
                    $L_{last\_activated}=U_{i}+1$;\\
                    $l_{i}$ ++;\\
                    $l_{last\_activated}=l_{i}$;\\
                }
                \If{last\_activated==M-1}
                {
                \For{i=0:M-1}{
                $l_{i}=L-1$;
                }}}}

\caption{CAT structure per memory bank}\label{Tree_algorithm1}
\end{algorithm}
\setlength{\textfloatsep}{0pt}

To determine whether to inspect the left or right entry, we examine the address $A$ ($0 \le A < N$). Starting at the root, the high order bit of $A$ determines the successor that covers row $A$, thus accessing a leaf or an intermediate nodes as already described.
More generally, when traversing an intermediate node at level $l$ of the CAT, the $l^{th}$ bit of $A$, counting from the most significant bit, is used to select the successor, which may be a leaf node or an intermediate node at level $l+1$. Before the CAT is completely built, it is guaranteed that fewer that $M$ counters and $M-1$ intermediate nodes are utilized.

To illustrate the process of splitting a counter during the building of the CAT, we consider the example CAT articulated in Figure~\ref{fig:table_tree} by rolling back the last split operation.  The last counter, C7 was deployed by splitting C6 into C6 and C7 and introducing I6.  Let us assume that the current I6 still points to a leaf node C6. 
At this point of the CAT construction, only 7 counters were deployed. This incomplete tree is represented by the same array $I$ of Figure~\ref{fig:table_tree} with the fourth entry of the array being [C3, \textbf{C6}, 0, \textbf{0}] (differences noted in bold) and the last entry being still undefined.  To reach the state shown in Figure~\ref{fig:table_tree}, C6 reached split threshold; I3's $R\_ptr$ was replaced by a pointer to the next available entry in $I$, I6; the last available counter, C7, is initialized to match C6; and I6 is set to [C6, C7, 0, 0] as is shown in the figure.  

Given the above implementation, the maximum number of sequential SRAM accesses for traversing the CAT is equal to the maximum depth of the tree, $L$. That number of accesses may be reduced if instead of starting to build the CAT tree from its root, we start from a pre-set complete binary tree with $\lambda$ levels for some $\lambda \le log M$. Consequently, to traverse the CAT, we can use the  most significant $\lambda$ bits of the row address, $A$, to directly access the appropriate intermediate node at level $\lambda - 1$, which reduces the maximum number of SRAM accesses to reach a leaf to $L - \lambda + 1$.  For example, if we start from a uniform binary tree with $\lambda = log M$ levels, the initial CAT will be a complete tree containing $M/2$ counters and $M/2 -1$ intermediate nodes. The other $M/2$ counters can then be used to grow the CAT non-uniformly beyond $\lambda$ levels and up to a maximum of $L$ levels. Moreover, by pre-splitting the counters uniformly up to level $\lambda - 1$ (that is starting from a balanced CAT with $\lambda$ levels), we can reduce the size of the intermediate node array because we can avoid storing the intermediate nodes at levels smaller than $\lambda$.
\begin{figure}[!t]
\begin{center}
  \includegraphics[width=0.475 \textwidth,height=0.145\textheight]{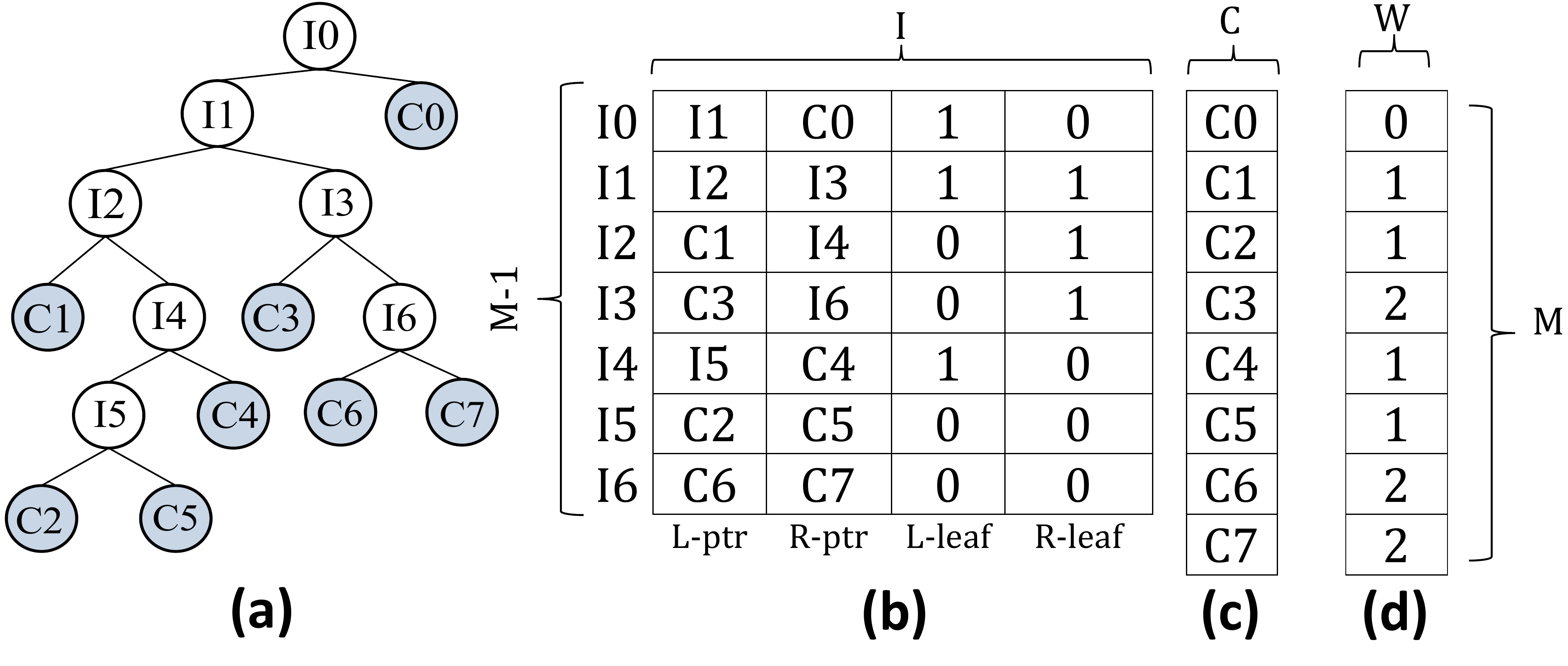}\vspace{-0.75em}
\caption{(a) A CAT using pointer chasing with $M=8$ counters and 6 levels. (b,c) The data structure used to represent the CAT. (d) An array of weight registers used for reconfiguring the CAT (see section 5).
}
\label{fig:table_tree}
\end{center}\vspace{-0.5em}
\end{figure}%

\subsection{Determining Split Threshold Values}

The CAT adapts the distribution of the available counters to the rows in a bank depending on memory reference patterns. Specifically, the CAT is dynamically shaped to minimize the number of refreshed rows, and thus, the refresh power. Given a sequence of row references, the split thresholds determine the shape of the tree. 
In experimenting with the CAT technique, we found that its performance is sensitive to the values of the split thresholds. Given the combinatorial number of options for selecting the split thresholds, we present in this section a model to determine these thresholds in a way that minimizes the number of refreshed rows. We explain that model assuming that we start from a uniform CAT with $\lambda=m=log M$ levels, and determine the split thresholds $T_{m-1}, ... , T_{L-2}$ used to grow the tree non-uniformly to a maximum of $L$ levels.

For illustration, we consider a simple example in which 4 counters are used for the $N$ rows in a bank. Specifically, assume that after a number of references, the CAT is represented by the tree structure shown in Figure \ref{fig:simple}(a). Depending on the reference pattern and the values of the split thresholds $T_{1}$ and $T_{2}$, this structure can evolve to either the balanced tree structure of Figure \ref{fig:simple}(b) or the non-balanced tree structure of Figure \ref{fig:simple}(c). That is, whether counter C1 splits first (Figure \ref{fig:simple}(b) or counter C3 splits first (Figure \ref{fig:simple}(c)) depends on the relative value of $T_{1}$ and $T_{2}$. After one of the counters splits, the CAT reaches its final shape and the thresholds of all the counters are set to the refresh threshold $T$. Hence, it is crucial to choose the split thresholds so that the CAT assumes the form of the tree that minimizes the number of refreshed rows.

Continuing with the 4-counter example, we note that a counter at level 0, 1, 2 and 3 will be assigned $4w$, $2w$, $w$ and $w/2$ rows, respectively, where $w = N/4$. Now, assume that the bank receives $R$ row references (accesses) during the regular refresh interval. 
If the references are uniformly distributed across rows, then each counter in Figure \ref{fig:simple}(b) will receive $R/4$ references, and if the refresh threshold is $T$, then each counter will reach this threshold $R/4T$ times. Each time a counter reaches $T$, the $w$ rows assigned to it are refreshed. Hence the total number of refreshes is\vspace{-0.5em}

\begin{equation}\label{rami1}
Cost_{SCA} = w \times R/T
\end{equation}

\begin{figure}[!t]
\begin{center}\vspace{0.75em}
 \includegraphics[width=0.465\textwidth]{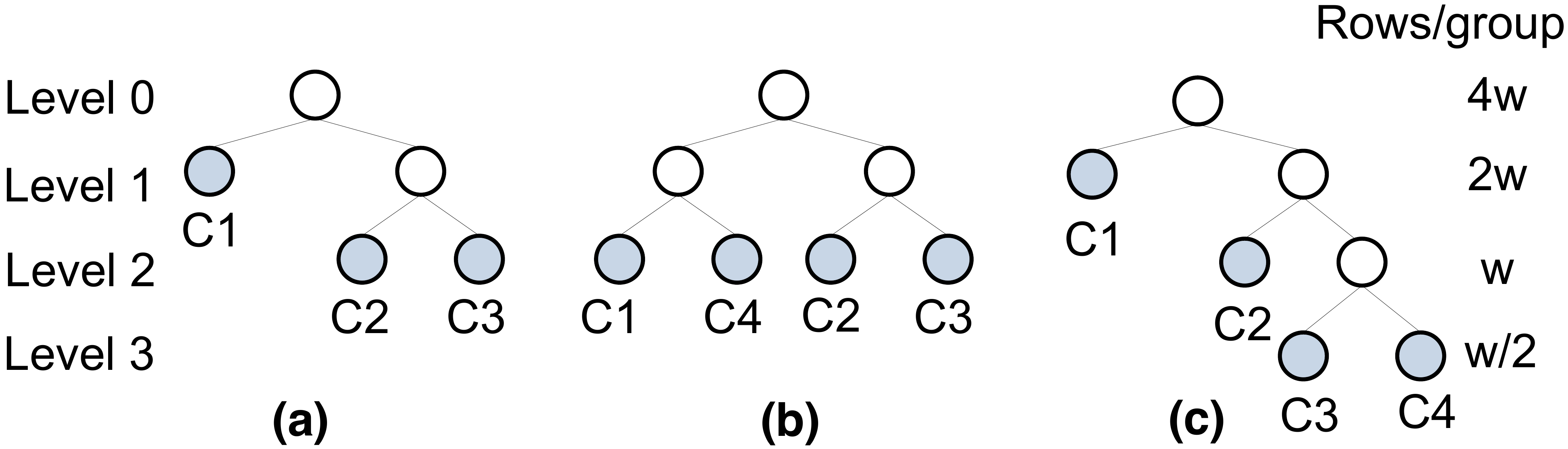}
\caption{Two possible evolutions of the CAT of (a) to a balanced tree structure (b) or an unbalanced structure (c). 
}\label{fig:simple}
\end{center}\vspace{-0.75em}
\end{figure}

An unbalanced CAT is expected to reduce the number of refreshed rows if the $R$ references are sufficiently biased towards a small group of rows. To determine the ``amount'' of bias that will favor the CAT of Figure~\ref{fig:simple}(c) over the uniform tree of Figure~\ref{fig:simple}(b), we define this bias using a variable $x$ such that the group of rows assigned to counter C4 receives $x$ more references than the other rows. That is the ratio of references caught by counters C1, C2, C3 and C4 is $2w:w:w/2:x+w/2$. This means that each of the four counters will receive $r1=2w \alpha$, $r2=w \alpha$, $r3=0.5 w \alpha$ and $r4=(x+0.5w) \alpha$ references, respectively, where $\alpha = R/ (x+4w)$. Consequently, the refresh threshold $T$ will be reached in counter C1 $r1/T$ times, and in each time the $2w$ rows assigned to it will be refreshed. Similarly, $T$ in C2 will be reached $r2/T$ times, and in each time the $w$ rows assigned to it will be refreshed. Finally, $T$ in C3 and C4 will be reached $r3/T$ and $r4/T$ times, respectively, and in each time the corresponding $w/2$ rows will be refreshed. Hence, the total number of refreshed rows is

\begin{equation} \label{rami2}
Cost_{CAT} = ((2w)^2 + w^2 + (\frac{w}{2})^2 + (x+\frac{w}{2}) \frac{w}{2}) \frac{\alpha}{T}
\end{equation}

From  (\ref{rami1}) and (\ref{rami2}), we conclude that $Cost_{CAT} < Cost_{SCA}$ when\vspace{-0.5em}
\begin{equation} \label{rami3}
x > 3w.
\end{equation}

After determining the critical bias that causes the CAT to outperform the uniform tree, we proceed to find the thresholds $T_{1}$ and $T_{2}$ that will force, after a short sequence of accesses, say $R_{s}$, the tree of Figure~\ref{fig:simple}(a) to evolve to the tree of Figure~\ref{fig:simple}(c) if $x > 3w$ and to the uniform tree otherwise. For this, we note that if the reference bias is $x = 3w$, then after $R_{s}$ references, the counters C1, C2 and C3 in Figure \ref{fig:simple}(a) will record $2w \beta$, $w \beta$ and $4w \beta$ accesses, respectively, where $\beta = R_s / 7w$. Hence, if $T_{2}$ is set to be $2 T_{1}$, then C3 will reach $T_{2}$ before C1 reaches $T_{1}$ when $x > 3w$, thus converging to the CAT of Figure \ref{fig:simple}(c). On the other hand, if $x < 3w$, then C1 will reach $T_{1}$ before C3 reaches $T_{2}$, thus leading to the uniform tree of Figure \ref{fig:simple}(b). To completely specify the split thresholds, we chose $T_{2} =T/2$, which guarantees that the CAT converges before any counter reaches the threshold $T$. Consequently, $T_{1} =T/4$.

The same reasoning used in the 4-counter example can be generalized to the case of $M = 2^m$ counters. Due to space limitation, this extension is not presented in this paper and is provided in the technical report. The generalized model is used to determine the split thresholds for all the experiments presented in this paper. For example, when applied to the tree with $M=64$ counters and $L=10$ levels, the values of the thresholds computed by the model are: $T_5$ = 5155, $T_6$ = 10309, $T_7$ = 12886, $T_8$ = 16384, and $T_9$ = $T$ = 32768.

\section{\textbf{RECONFIGURING THE CAT TO TRACK CHANGES IN ACCESS PATTERNS}}

The CAT assigns the available counters to the rows of a bank according to the pattern of row accesses. However, the row access pattern changes with time, which necessitates a mechanism for the reconfiguration of the CAT to track these changes. In the next two sections, we propose two such mechanisms. The first, PRCAT, periodically reconstructs the CAT and the second, DRCAT, dynamically reconfigures the CAT by reassigning counters from cold to hotter regions of the bank.

\subsection{Periodically Reset CAT (PRCAT)}

In this scheme, the CAT tree is rebuilt at epochs equal to the auto-refresh interval (64ms for several DRAM generations ~\cite{chatterjee2012usimm}). For LPDDRx devices that support burst refresh~\cite{bhati2016dram}, this simple scheme tracks the number of row accesses exactly. It can also be applied to modern DDRx devices that support distributed refresh at the expense of some inaccuracy in tracking the number of accesses. Specifically, because row refreshes are out of sync with the resetting of the CAT, recent information about row accesses are lost when the CAT is reset. Moreover, PRCAT resets the CAT periodically, even when the row access patterns do not change, potentially incurring the overhead of reconstructing the CAT unnecessarily. In the next section, we describe a CAT reconfiguration scheme which avoids these two shortcoming at a small cost for keeping additional information about the usage of the counters in a CAT.
\vspace{-0.15em}\subsection{Dynamically Reconfigured CAT (DRCAT)}\label{DRCAT}
The DRCAT allocates weights to counters to track the number of times each counter reaches the refresh threshold. After the CAT is completely built, the DRCAT identifies the counters allocated to regions that become cold and reallocate them to regions that become hot. A 2-bit weight register is used to record the weight of each counter. As described in the last section, when a counter reaches the refresh threshold, its corresponding rows are refreshed and its value is reset to zero. However, to keep track of the hotness of row regions, the weight corresponding to that counter is incremented (with an upper bound of 3) and the weights corresponding to all other counters are decremented (with a lower bound of 0). If the weight of the incremented counter reaches its maximum limit, two counters having zero weights (cold regions) are merged and the released counter is used to split the hot counter.

To illustrate the scheme, consider the CAT example shown in Figure~\ref{fig:table_tree}, where all counters have been activated and the weights of the counters are kept in the register $W$ depicted in Figure~\ref{fig:table_tree}(d). Assume that at a given time during operation, the values of the weight registers are [0,1,1,2,1,1,2,2] and counter C6 reaches its limit (we used 2-bit counters). After the rows corresponding to C6 are refreshed, the values of the weights are updated to [0,0,0,1,0,0,3,1] and the following steps are taken to reconfigure the CAT:\\
(1) From the table shown in Figure~\ref{fig:table_tree}(b) an intermediate node in the CAT which has two counters as children (L\_leaf = R\_leaf = 0) with the weight of both being zero is selected. If such a node is found, the two counters are merged, one counter is freed and we go to step 2.  In our example, C2 and C5 are leaves and both weight registers are zero.  Hence, C5 is promoted to its parent node and the fifth row of the table is updated to I4=[\textbf{C5},C4,\textbf{0},0] as shown in Figure~\ref{fig:rcbt1}(b). Furthermore, C2 and the sixth row, I5, of the table are released.\\ 
(2) We split the region tracked by the hot counter using the counter freed in step 1.  In our example, we show splitting C6 by replacing the L\_ptr in its parent node (entry I6) by the index of the released row (I5) and set its corresponding flag to 1 to indicate that I5 will represent an intermediate node. Finally, we update I5=[\textbf{C6,C2,0,0}] to point to C6 and C2 and reset the corresponding flags. \\
(3) We update weight of the newly split counters to 1 to ensure they remain split for a reasonable period of time while preventing them from being quickly split in succession.

The DRCAT adds a negligible area overhead to the PRCAT design. For example, PRCAT uses 2 bytes per counter for T=16K and in this case, it occupies area overhead similar to DRCAT. The reason is that DRCAT uses the first 16 bits for the counter and the two last bits for the weight register.
With respect to latency, the DRCAT traverses the tree to find the cold counters and their parent intermediate node. Since the reconfiguration of the tree happens infrequently and traversing the tree is not on the critical path, system's performance is not affected by the reconfiguration. 

Note that, in addition to tracking the change in the hot spot of memory accesses, the reconfiguration of the CAT according to the weights of the counters has the flexibility of adapting to multiple hot spots in the access patterns.
\begin{figure}[!t]
\begin{center}
 \includegraphics[width=0.475 \textwidth,height=0.15\textheight]{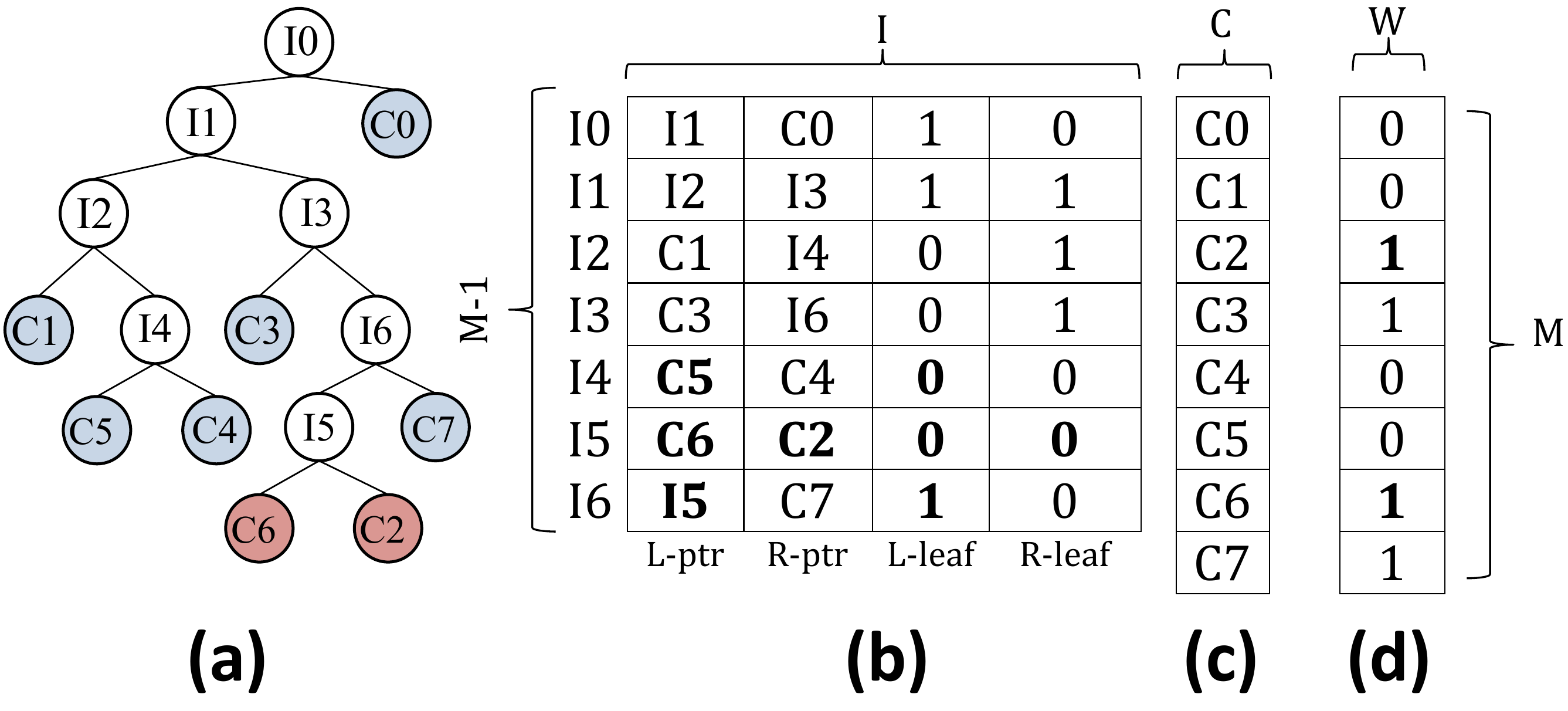}\vspace{-0.75em}
\caption{The CAT of Figure \ref{fig:table_tree} after reconfiguration.}
\label{fig:rcbt1}
\end{center}
\end{figure}%

\section{\textbf{EXPERIMENTAL METHODOLOGY}}\label{Methodology}

To evaluate the proposed technique, we performed simulations using the memory system simulator USIMM modeling 55nm DRAM~\cite{chatterjee2012usimm}. Unless stated otherwise (in Section~\ref{sensitivity}), the default simulation environment was set to model memory traffic from a dual core CPU. The total memory capacity is 16 GB with a total of 16 banks divided into two ranks, with 64K rows per bank. The Last level cache size is 512KB per core in our simulation. Detailed simulation parameters for USIMM are listed in Table~\ref{tab:setting}. The DRAM timing constraints follow a Micron DDR3 SDRAM data sheet~\cite{DDR3,jacob2010memory}.
Verilog implementations of the control logic for the different wordline crosstalk mitigation schemes were created to provide an area and energy overhead comparison. These Verilog codes were synthesized using Synopsys Design Compiler and evaluated for power using Synopsys PrimeTime, targeting a 45nm FreePDK standard cell library~\cite{FreePDK45,seyedzadeh2018enabling,seyedzadeh2017mitigating,seyedzadeh2016leveraging,seyedzadeh2016improving,seyedzadeh2015pres}\footnote{It is commonplace for DRAM to trail CMOS by a technology generation. Systems with 45nm CPUs were concurrent with 55nm DRAM.}. 
We have changed the number of counters per bank in the designs between 32 and 512 and, for CAT, allowed the trees to grow up to 14 levels to study the trade-off between performance, crosstalk mitigation refresh power, and hardware overhead. 
For the evaluation of PRA, we accounted for the energy to generate a random number every row access. 

To provide realistic workloads for evaluating the wordline crosstalk mitigation schemes, we used workloads from the Memory Scheduling Championship~\cite{msc}. These workloads cover a variety of benchmarks including commercial applications and selected benchmarks from the PARSEC, SPEC, and Biobench suites. Furthermore, we use 12 kernel attacks to mimic malicious codes in Section \ref{malicious}.

One metric used to compare different crosstalk mitigation schemes is the crosstalk mitigation refresh power overhead (CMRPO). The CMRPO is the average power consumed for deciding which rows to be refreshed in order to mitigate crosstalk. It is computed relative to the regular refresh power in the absence of any crosstalk mitigation (2.5mW to refresh 64K rows during a 64 $ms$ refresh interval~\cite{DDR3,rahmati2014refreshing}).

While rows are refreshed in a bank to mitigate crosstalk, that bank cannot be accessed, possibly delaying subsequent memory requests to that bank. To estimate this delay, we define the execution time overhead (ETO) as the delay in execution time due to memory requests to banks being refreshed (to mitigate crosstalk) relative to the execution time when no provisions are made to mitigate crosstalk.
\renewcommand{\arraystretch}{1.5}
\begin{table}[!htb]
\footnotesize 
\centering
\caption{System Configuration}\vspace{-0.9em}
\label{tab:setting}
\scalebox{0.975}{
\begin{tabular}{|l|l|}
\hline
\small Processor                                                    & \begin{tabular}[c]{@{}l@{}}\small Two 3.2GHz cores, Memory bus speed: 800 MHz\\ 128-entry ROB, Fetch width: 4, Retire width: 2\\ Pipeline depth: 10\end{tabular}                                           \\ \hline
\begin{tabular}[c]{@{}l@{}}\small Memory \\ \small controller\end{tabular} & \begin{tabular}[c]{@{}l@{}}\small Bus freq.: 800 MHz,Write queue capacity: 64\\ Address mapping: rw:rk:bk:ch:col:offset\\ Management policy: closed-page with FRFCFS\end{tabular} \\ \hline
\small DRAM                                                         & \begin{tabular}[c]{@{}l@{}}\small 2 channels(each 8GB DIMM), 1 rank/channel\\ 8 banks/rank, 64K rows/bank, 64B cache line\end{tabular}                                            \\ \hline
\end{tabular}
}\end{table}
\renewcommand{\arraystretch}{1.3}
\begin{table*}[!t]
\centering
\caption{Hardware energy (per bank) and area of DRCAT, PRCAT and SCA for different number of counters, $M$, and the specification of the PRNG used for PRA~\cite{srinivasan20102}. The reported energy for PRNG (\textit{eng\_PRNG}) is for generating 9-bits per row access.
}
\label{tab:2}\vspace{-0.7em}
\scalebox{0.76}{
\begin{tabular}{|l|l|l|l|l|l|l|l|l|l|l|l|}
\hline
\multirow{3}{*}{M} & \multicolumn{6}{l|}{Energy:dynamic (nJ per row access) and static (nJ per refresh interval)} & \multicolumn{3}{l|}{\;\;\;\;\;\;\;\;\;\;\;\;\;Area ($mm^{2}$)}                                     & \multicolumn{2}{l|}{}                                                \\ \cline{2-10} 
                   & \multicolumn{2}{l|}{\;\;\;\;\;\;\;\;\;\;\;DRCAT}   & \multicolumn{2}{l|}{\;\;\;\;\;\;\;\;\;\;PRCAT}  & \multicolumn{2}{l|}{\;\;\;\;\;\;\;\;\;SCA}  & \multirow{2}{*}{DRCAT} & \multirow{2}{*}{\;\;\;\;PRCAT} & \multirow{2}{*}{\;\;\;\;SCA} & \multicolumn{2}{l|}{\multirow{2}{*}{\;\;\;\;\;\;\;\;\;\;\;\;\;\;\;\;\;\;\;PRNG}} \\ \cline{2-7}
                   & \;\;dynamic          & \;\;\;\;static          & \;\;dynamic         & \;\;\;\;static         & \;\;dynamic         & \;\;\;\;static         &                       &                      &                      & \multicolumn{2}{l|}{}                                                   \\ \hline
32             & \;3.05E-04	&\;5.77E+03\;& \;2.91E-04\;& \;5.55E+03\;& \; 1.41E-04 &\;3.16E+03 &  3.16E-02 &  3.04E-02 & 1.86E-02 & Area              &\;\; 4.004E-3\\ \hline
64             & \;4.30E-04	&\;1.39E+04\;& \;4.09E-04\;& \;1.32E+04\;& \; 1.92E-04 &\;8.81E+03 &  6.12E-02 &  5.86E-02 & 4.04E-02 & Throughput(Gbps)  &\;\;\;\; 2.4\\ \hline
128            & \;5.83E-04	&\;2.77E+04\;& \;5.50E-04\;& \;2.63E+04\;& \; 2.22E-04 &\;1.44E+04 &  1.16E-01 &  1.11E-01 & 6.04E-02 & Power(mW)&\;\;\;\;\; 7  \\ \hline
256            & \;8.72E-04	&\;5.44E+04\;& \;8.25E-04\;& \;5.13E+04\;& \; 3.12E-04 &\;2.39E+04 &  2.23E-01 &  2.11E-01 & 1.00E-01 & Eff.(nj/b)        &\;\; 2.90E-3 \\ \hline
512            & \;1.17E-03	&\;1.06E+05\;& \;1.10E-03\;& \;1.02E+05\;& \; 4.25E-04 &\;4.52E+04 &  3.93E-01 &  3.75E-01 & 1.72E-01 & eng\_PRNG(nj)  &\;\; 2.625E-2                       \\ \hline
\end{tabular}
}
\end{table*}
\section{\textbf{EVALUATION}}\label{Evaluation}

We compare crosstalk mitigation schemes: DRCAT, PRCAT, SCA (implemented with SRAM) and PRA (refreshes two victim rows but not the aggressor row). In this section, we conduct experiments on a dual-core system using refresh thresholds of T=32K and T=16K and a maximum of L=11 levels for DRCAT and PRCAT.
In Section~\ref{CMRPO_Sensitivity_cbt_th}, we will study the effect of the maximum number of CAT levels and the value of the refresh thresholds on power and performance. Moreover, we will report results for quad-core systems. We assume that either the memory controller knows which rows are physically adjacent to each other~\cite{van2002address} or the DRAM chip is responsible for refreshing the row and its neighbors~\cite{bains2016row}.

\subsection{Hardware Overhead}\label{hardware}

Table~\ref{tab:2} shows the hardware cost for managing and maintaining the counters for SCA, DRCAT and PRCAT with $L$=11 levels and $T=32K$ as the number of counters per bank ranges from 32 to 512. We separately report the dominant sources of hardware energy overhead. These sources include: (1) the dynamic energy per access of the designed circuits plus the SRAM storage, and (2) the static energy during a 64 $ms$ refresh interval of circuits plus the SRAM storage. The SRAM energy is extracted from CACTI~\cite{muralimanohar2009cacti} and the circuit energy (combinational and io-pad) is derived from Synopsys PrimeTime. Note that for DRCAT and PRCAT, the dynamic energy per memory access accounts for multiple accesses to SRAM (from 2 to $L-log(M/4)$) while for SCA, SRAM is accessed only twice to read and write the counters. A modified version of Table~\ref{tab:2} is used for DRCAT and PRCAT when the maximum tree depth changes in the experimental tests.

The results show that the dynamic energy per access of PRCAT is roughly twice that of SCA for the same number of counters. With respect to area overhead and static energy, Table~\ref{tab:2} clearly shows that PRCAT and SCA occupy equal area and consume similar static power when the number of counters of SCA is twice that of PRCAT. For example, PRCAT$_{64}$ and SCA$_{128}$ occupy iso-area. Moreover, this area is one order of magnitude smaller than the area needed by the leading counter-based approach that stores in memory one counter per row and uses a 32KB on-chip counter cache~\cite{kim2015architectural} (equivalent storage to 2,048 counters per bank). Thus, implementing 64 or even 256 counters per bank is feasible. Our implementation shows that the average latency for PRCAT is 3.6$ns$ (circuit latency plus repeated access to SRAM) which is much lower than the row activation latency in the DRAM memory~\cite{shin2016dram}.

In comparison to PRCAT for T=32K, DRCAT uses a 2-bit weight register per counter to reconfigure the structure of CAT. The results in Table~\ref{tab:2} show that the circuit design and SRAM storage of DRCAT, on average, augments 4.2\% area overhead to the system compared to PRCAT. Also, DRCAT increases the dynamic energy per row access by 5\% over PRCAT. Furthermore, it incurs 4ns latency. When DRCAT reconfigures counters, its latency is about 7.5ns. The main reason for the extra latency is the traversal of the tree as explained in Section~\ref{DRCAT}. However, updating the DRCAT and accessing the memory can be done in parallel.

Table~\ref{tab:2} also shows the specification of a PRNG~\cite{srinivasan20102} for PRA in 45nm technology\footnote{ An PRNG design with low static power is reported in \cite{yang201416}. However, this design is much slower than the design in \cite{srinivasan20102} which leads to a larger Energy/bit consumption.}. 
We select one PRNG for PRA that is applied for all banks during row accesses.
The energy per bit (the efficiency) for PRNG is computed as \textit{Power}/\textit{Throughput}. For $p=0.002$ and $p=0.003$, PRNG generates 9 bits ($\sim log({1/0.003}$ or $log({1/0.002})$) so that PRA can decide if victim rows should be refreshed when a row is accessed. The energy for generating 9 random bits is denoted by $eng\_PRNG$. A similar conclusion was reached in \cite{armor}.

\subsection{CMPRO}\label{total_energy}

We use the results shown in Table~\ref{tab:2} to compute CMRPO for a benchmark during its execution by adding the following components needed to mitigate crosstalk: (1) The dynamic power (product of dynamic energy per memory access and the total number of memory accesses during execution divided by the execution time), (2) the static power (static energy during a refresh interval divided by the refresh interval), and (3) the refresh power (product of the average number of rows refreshed to prevent crosstalk with the energy to refresh one row (1nJ per row~\cite{ghosh2007smart}) divided by the execution time).

Figure~\ref{Ex-fig1} shows the CMRPO for different approaches when $T=32K$. It reveals that both DRCAT$_{64}$ and PRCAT$_{64}$ with L=11 achieve a CMRPO of 4\%, which is an improvement over the 11\% in the cases of PRA and SCA. Note that the CMRPO for PRA includes refreshing an average of two victim rows every 500 accesses and generating 9 PRNG bits every access, with the PRNG generation being dominant. According to Table~\ref{tab:2}, on average, for every 50 row accesses, PRA consumes energy equal to that of refreshing one row in DRAM.

For T=16K, we use PRA$_{0.003}$, rather than PRA$_{0.002}$ since the probability of failure for PRA$_{0.002}$ is greater than 1E-4 (Chipkill reliability) according to Figure~\ref{pra_reliability}. Figure~\ref{Ex-fig1} shows that CMRPO for DRCAT$_{64}$ in dual-core systems is 4.5\%, which is an improvement over the 12\% and 22\% incurred in PRA$_{0.003}$ and SCA$_{64}$, respectively. Also, considering iso-area, DRCAT$_{64}$ achieves a CMRPO improvement over the 13\% incurred in SCA$_{128}$. Figure~\ref{Ex-fig1} indicates that reducing $T$ from 32K to 16K will increase considerably CMRPO for SCA while slightly increasing CMRPO for PRCAT and DRCAT.\vspace{-0.25em}
\begin{figure*}[!t]
\begin{center}
\includegraphics[width=1.02\textwidth]{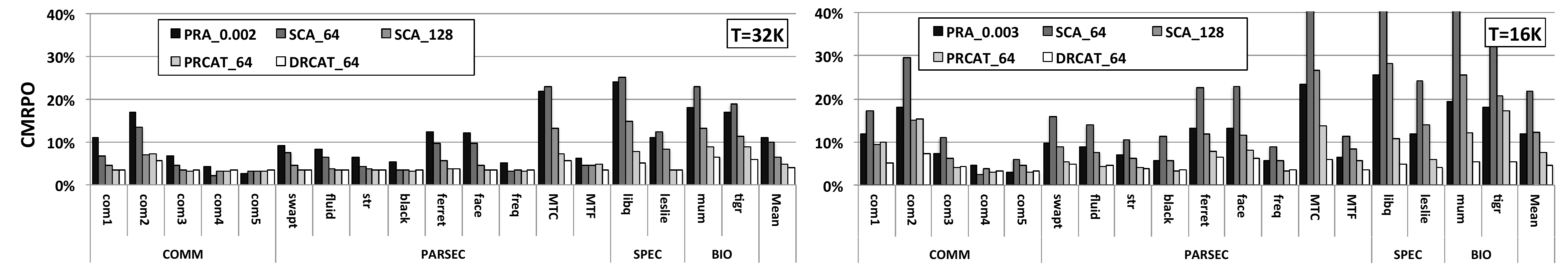}\vspace{-0.25em}
\caption{The CMRPO  (as a percent of the regular refresh power). DRCAT and PRCAT use 64 counters and up to 11 levels.}\label{Ex-fig1}
\end{center}\vspace{-1.05em}
\end{figure*}
\begin{figure*}[!t]
\begin{center}
\includegraphics[width=1.02\textwidth]{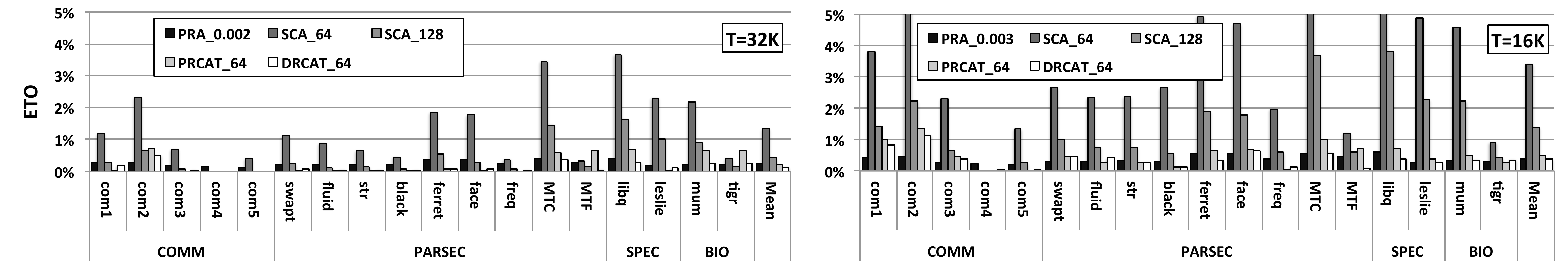}\vspace{-0.5em}
\caption{ETO resulting from refreshing vulnerable rows. DRCAT and PRCAT use 64 counters and up to 11 levels.
}\label{Ex-fig7}
\end{center}\vspace{-1.75em}
\end{figure*}
\subsection{Execution Time Overhead}\label{performance}

To evaluate performance, we report the execution time overhead (ETO) resulting from refreshing victim rows. When rows vulnerable to crosstalk are refreshed, any read or write request to the bank containing the refreshed rows is stalled, which leads to the execution time overhead.

Figure~\ref{Ex-fig7} shows the ETO for different workloads. For $T=32K$, PRA$_{0.002}$, SCA$_{64}$, SCA$_{128}$, PRCAT$_{64}$ and DRCAT$_{64}$ incur low ETO of 0.26\%, 1.32\%, 0.43\%, 0.23\%, and 0.16\% respectively. For $T=16K$, the ETOs of PRA$_{0.003}$, SCA$_{64}$, SCA$_{128}$, PRCAT$_{64}$ and DRCAT$_{64}$ are 0.39\%, 3.42\%, 1.38\%, 0.49\% and 0.35\% respectively. Note that ETO for PRA$_{0.003}$ when $T=16K$ is roughly 1.5 times larger than ETO for PRA$_{0.002}$ when $T=32K$ because it probabilistically refreshes 50\% more rows. On the other hand, ETO for SCA$_{128}$  when $T=16K$ is higher than ETO for SCA$_{64}$  when $T=32K$. This shows that when the refresh threshold is reduced, doubling the number of counters statically does not reduce the number of refreshed rows, which results in less accurate row tracking and thus larger refresh energy.
\begin{figure*}[!t]
\begin{center}
\includegraphics[width=0.92\textwidth]{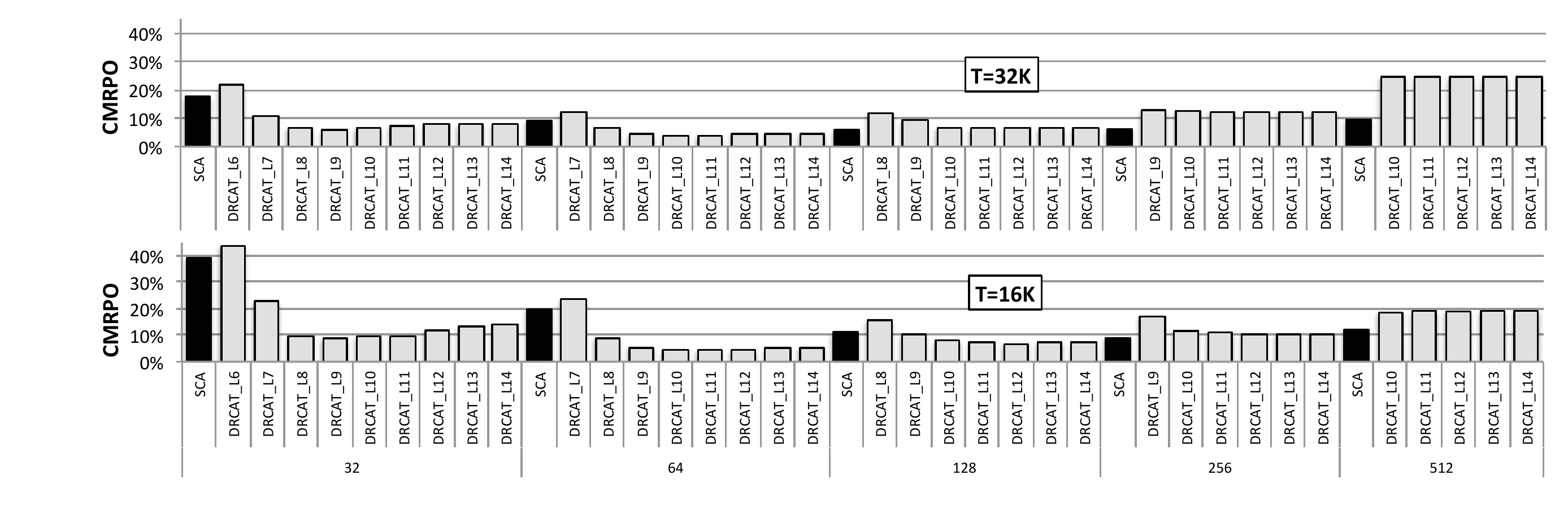}\vspace{-0.5em}
\caption{Crosstalk mitigation power overhead per bank for DRCAT using from 32 to 512 counters and different maximum CAT levels (6 to 14).
}\label{Ex-fig10x}
\end{center}\vspace{-1.5em}
\end{figure*}
\begin{figure*}[!t]
\begin{center}
\includegraphics[width=0.92\textwidth]{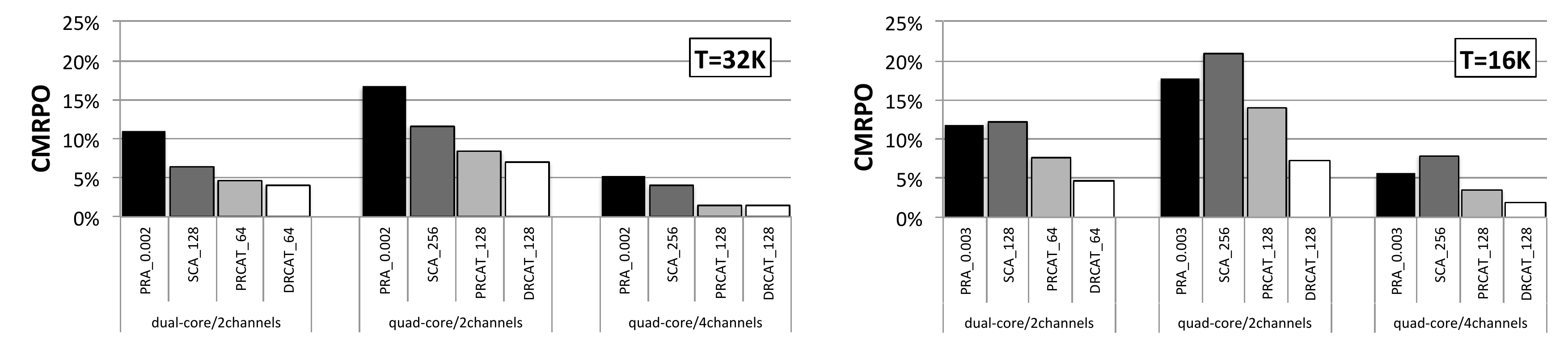}
\caption{Effect of different mapping polices and number of cores on CMRPO (per bank). The banks in dual core and quad core systems include 64K and 128K rows, respectively.}
\vspace{-0.5em}
\label{mp_cmpo}
\end{center}\vspace{-1.92em}
\end{figure*}

\section{\textbf{SENSITIVITY STUDY}}\label{sensitivity}

\subsection{Sensitivity to the Number of Counters and the Maximum CAT depth}\label{CMRPO_Sensitivity_cbt_th}
Figure~\ref{Ex-fig10x} shows CMRPO for DRCAT when the number of counters changes from 32 to 512 and the number of levels changes from 6 to 14, and compares results with those of SCA. From the figure, we note that increasing the number of CAT levels does not significantly impact CMRPO when the number of counters is relatively large. This is because, in this case, the static power consumed by the counters dominates the CMRPO, and hence, any improvement in the number of refreshed rows has minimal effect. Conversely, with a small number of counters, the energy for refreshing vulnerable rows is a large component of the CMRPO. Thus, having more levels in the tree saves refresh energy by targeting vulnerable row.

Due to the trade-off between static power and the power consumed to refresh vulnerable rows, the minimum CMRPO happens when DRCAT employs 64 counters and when SCA employs 128 counters for T=32K. Note that the refresh power of DRCAT$_{64}$ with L7 is close to SCA$_{64}$ since it only increases row resolution one more level beyond SCA$_{64}$. However, DRCAT$_{64}$ incurs more static and dynamic power than SCA$_{64}$; hence, its CMRPO is larger. The same argument applies to explain why for fewer counters, CMRPO of SCA$_{32}$ is smaller than that of DRCAT$_{32}$. When the threshold decreases from 32K to 16K, SCA will refresh victim rows more frequently and its CPRMO grows by 12\% while the minimum CMRPO of DRCAT$_{64}$ changes very little.

We studied the sensitivity of ETO to the number of counters and the tree depth (the results are not shown in this paper). The key observation is that, for both refresh thresholds, when using at least 64 counters and $L \geq 9$, DRCAT incurs an ETO < 1\%. Results also show that ETO is inversely correlated to the refresh threshold. Another observation is that for a given fixed number of counters, increasing the tree depth does not necessarily reduce the number of refreshed victim rows; with a deeper tree, the number of rows associated with a certain counter will be reduced, but the number of rows associated with other counters will increase. In other words, trying to be precise in one area of the memory may lead to a gross imprecision in another area of the memory, which creates a trade-off that leads to an optimum value for the maximum tree depth.

We conclude that for DRCAT, the optimal number of counters and the maximum CAT depth affect both the CMRPO and the ETO. For $T=32K$ and $T=16K$ and using between 32 and 128 counters, a maximum of $L=11$ levels minimizes CMRPO and results in a low ETO. For CAT with more counters, the maximum CAT depth is inconsequential for CMRPO. In fact, using DRCAT leads to larger CMRPO than using SCA. We did the same analysis for PRCAT and our results show that CMRPO for PRCAT is about 4\% and 7\% for T=32K and T=16K with 10 and 11 CAT levels, respectively. Also it incurs very low performance overhead (<0.5\%) for both thresholds.

\subsection{Sensitivity to Mapping Policy and Number of Cores}

To analyze the effect of address interleaving, we experiment with dual-core systems using two standard mapping policies of USIMM~\cite{chatterjee2012usimm}: (1) the 2-channel mapping policy (used in the experiments so far) and (2) a 4-channel mapping policy that maximizes memory access parallelism. Note that when keeping the size of each memory bank fixed, the 4-channel policy in USIMM quadruples the number of banks in the system. We also experiment with a quad-core system using the 2-channel and 4-channel mapping policies. The CMRPO of DRCAT, PRCAT and SCA are reported in Figure~\ref{mp_cmpo} for iso-area storage.
Figure~\ref{mp_cmpo} shows that, when using the 2-channel mapping policy, the CMRPO for quad-core systems is larger than the dual-core systems. This is because having more cores reduces the spatial locality in the L2 cache, thus generating more memory traffic and forcing more refreshes. SCA is affected more than the other schemes by the increased traffic because of the inability to accurately track the row accesses due to the uniform distribution of counters to rows. This effect is amplified when $T=16K$ resulting in the CMRPO for SCA exceeding that of PRA for the quad-core system. In this case, DRCAT reduces the CMRPO in quad-core systems to 7\%, which is an improvement over the 21\% and 18\% incurred in SCA and PRA, respectively.
Figure~\ref{mp_cmpo} shows that for quad-core systems, the 4-channel policy reduces CMRPO versus the 2-channel policy for all schemes. This is expected since in the 4-channel policy, the number of banks increases from 16 to 64, thus decreasing the number of refreshed rows.

Although we do not show the results for ETO in this section, we should note that ETO remains low for all schemes irrespective of the mapping policy or the number of cores. The largest ETO is incurred when the 2-channel policy is used with quad-cores and $T=16K$. Specifically, in this case ETOs for PRA$_{0.003}$, SCA, PRCAT and DRCAT are 0.47\%, 1.45\%, 0.6\%, 0.38\% respectively. The relatively high ETO for SCA is due to the fact that the number of refreshed rows is relatively high.

\begin{figure}[!t]
\begin{center}\vspace{-0.75em}
\includegraphics[width=0.49\textwidth]{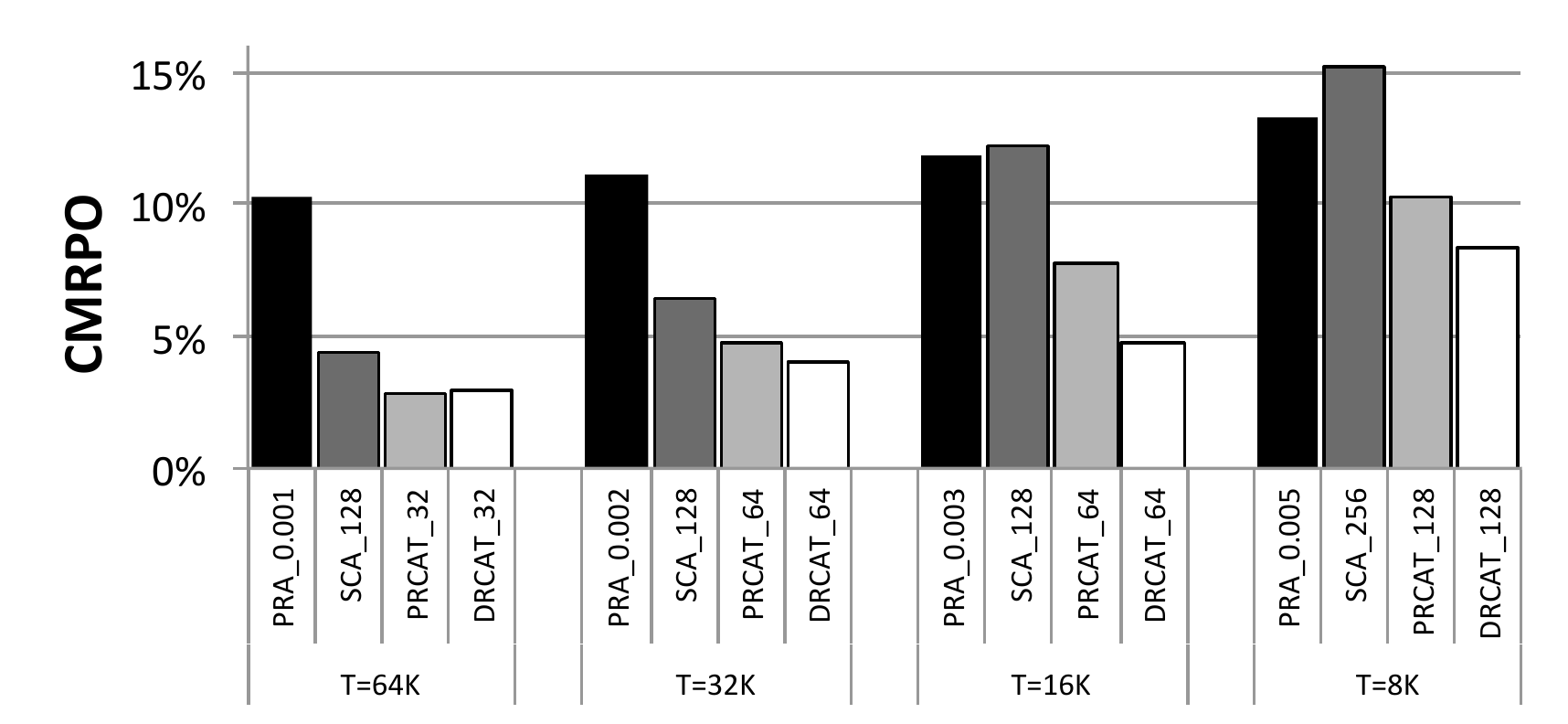}
\caption{CMRPO for refresh thresholds $T$ = 64K/32K/16K/8K.}
\label{Ex-fig12X}
\end{center}\vspace{-0.05em}
\end{figure}

\subsection{Sensitivity to Refresh Thresholds}\label{Sensitivity_Rfsh_th}

Scaling down DRAM technology exacerbates the crosstalk problem leading to a decrease in the refresh threshold~\cite{kim2015architectural}. This motivates the sensitivity analysis on different refresh thresholds presented in Figure~\ref{Ex-fig12X}, which shows the CMRPO for four refresh thresholds on a dual-core system with the 2-channel mapping policy. We used PRA$_{0.001}$, PRA$_{0.002}$, PRA$_{0.003}$ and PRA$_{0.005}$ for $T$ = 64K, 32K, 16K and 8K, respectively to guarantee that the unsurvivability is better than 1.0E-4.
The figure shows that, for thresholds 64K to 16K and dual core systems, DRCAT incurs CMRPO less than 5\% which is an improvement over PRA's 12\%. Also, it improves the CMRPO over PRCAT because the CAT is dynamically reconfigured rather than being periodically reset. Note that for T=8K, DRCAT and PRCAT need to double the number of counters to mitigate crosstalk, but still incur less than 10\% CMRPO. With respect to ETO, all approaches incur very low overhead. Specifically, for $T=8K$, the ETOs for PRA, SCA, PRCAT, DRCAT are 0.58\%, 1.44\%, 0.8\%, and 0.48\%, respectively. We conclude that CAT improves  CMRPO relative to the other schemes for both current and future technologies.

\subsection{Performance Under Malicious attacks}\label{malicious}

To evaluate the performance of the counter-based approaches under malicious attacks, we use 12 kernel attacks~\cite{armor} that randomly select few target rows~(4 rows per bank and a total of 64 target rows for 16 banks with dual-core/2-channels configuration) and access the target rows more frequently than other rows in DRAM. We integrate the kernel attacks with regular access rows of memory-intensive workloads (which we call benign workloads). We select three attack modes \textit{Heavy} (75\% target rows + 25\% benign access rows), \textit{Medium} (50\% target rows + 50\% benign access rows) and \textit{Light} (25\% target rows + 75\% benign access rows). Note that the distribution of target rows in the kernel attacks follows the Gaussian distribution. Figure~\ref{fig-kernel_attack} shows the average execution time overhead for the benign workloads for three refresh thresholds. 
As expected, more intensive attacks leads to higher ETO since it causes more refreshes.
While the ETO for PRCAT and DRCAT is less than 0.9\% and 0.6\% for different attacks and refresh thresholds, the ETO of SCA grows to 4.5\% for T=16K under heavy attacks. 
ETO for $T=8K$ is lower than for $T=16K$ because the number of counters is doubled.

We conclude that when malicious attacks target specific rows in DRAM, CAT-based approaches are more efficient than SCA approaches at mitigating the attacks since they confine attacked rows to smaller groups of rows to be refreshed.

\begin{figure}[!t]
\begin{center}\vspace{-0.75em}
\includegraphics[width=0.495\textwidth]{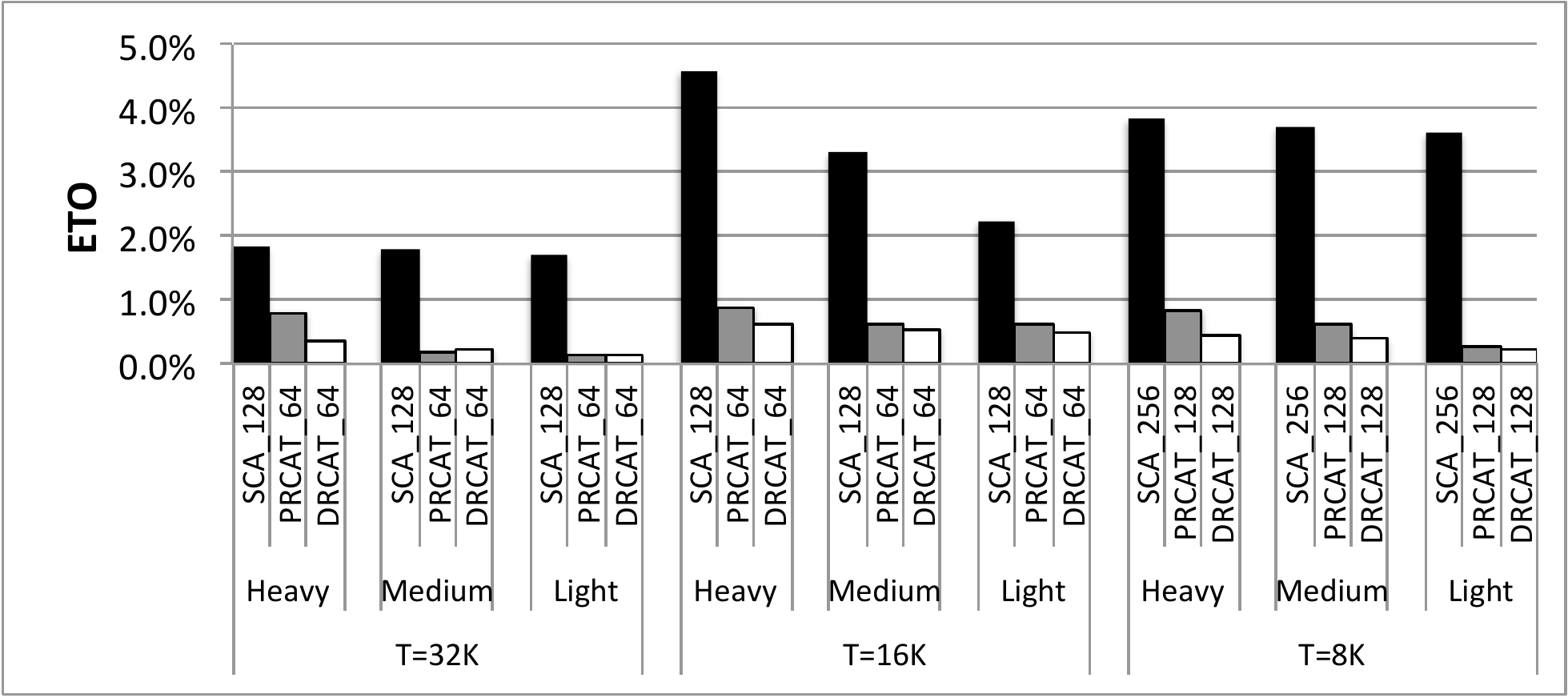}
\caption{ETO for three kernel attack modes: Heavy~(75\% target rows + 25\% benign access rows), Medium~(50\% target rows + 50\% benign access rows) and Light~(25\% target rows + 75\% benign access rows).}
\label{fig-kernel_attack}
\end{center}\vspace{-0.05em}
\end{figure}

\section{\textbf{CONCLUSION}}

We introduce the notion of a tree-based non-uniform row partitioning for detecting rows vulnerable to crosstalk in memory banks. We develop a low-cost implementation of this notion with three key ideas: 
(1) we propose a low-cost implementation to maintain and access Counter-based Adaptive Trees that assign counters to rows non-uniformly and detects more precisely rows vulnerable to crosstalk. (2) We introduce a scheme to compute the split thresholds that cause the trees to dynamically evolve and match the row access patterns. (3) We introduce a scheme, DRCAT, for dynamically reconfiguring the CAT to track the temporal changes in memory access patterns resulting from either changing the running applications or changing the phases of a running application.

Our results show that DRCAT outperforms the leading approaches for wordline crosstalk mitigation. Specifically, for quad-core systems and refresh threshold of $T=16K$, DRCAT reduces the CMRPO to 7\%, which is an improvement over the 21\% and 18\% incurred in deterministic and probabilistic approaches, respectively. Moreover, DRCAT incurs very low performance overhead ($< 0.5\%$). Hence, we conclude that dynamic row partitioning is an effective solution to detect rows vulnerable to crosstalk in DRAM. Clearly, this hardware solution avoids wordline crosstalk during normal execution and protects against malicious attacks that explore vulnerability to wordline crosstalk.

\section{\textbf{ACKNOWLEDGEMENTS}}

We thank the anonymous reviewers for their feedback. This work is supported by CS50 merit pre-doctoral fellowship award from the university of Pittsburgh.

\small
\setlength{\bibsep}{0pt plus 0.4ex}
\bibliographystyle{unsrt}
\bibliographystyle{reference}
\bibliography{reference}
\end{document}